\documentclass[preprint,prd,nofootinbib,tightenlines,amsmath]{revtex4}
\usepackage{enumerate}
\usepackage{multirow}

\usepackage{epsfig}
\usepackage{graphics,color}      % usual driver
\usepackage{verbatim}
\usepackage{amsmath}    % need for subequations
\catcode`\@=11
\def\sla@#1#2#3#4#5{{%
 \setbox\z@\hbox{$\m@th#4#5$}%
 \setbox\tw@\hbox{$\m@th#4#1$}%
 \dimen4\wd\ifdim\wd\z@<\wd\tw@\tw@\else\z@\fi
 \dimen@\ht\tw@
 \advance\dimen@-\dp\tw@ \advance\dimen@-\ht\z@
 \advance\dimen@\dp\z@
 \divide\dimen@\tw@ \advance\dimen@-#3\ht\tw@
 \advance\dimen@-#3\dp\tw@ \dimen@ii#2\wd\z@
 \raise-\dimen@\hbox to\dimen4{%
 \hss\kern\dimen@ii\box\tw@\kern-\dimen@ii\hss}%
 \llap{\hbox to\dimen4{\hss\box\z@\hss}}}}

\def\slashed#1{%
 \expandafter\ifx\csname sla@\string#1\endcsname\relax
{\mathpalette{\sla@/00}{#1}}
\fi}
\def\declareslashed#1#2#3#4#5{%
 \expandafter\def\csname sla@\string#5\endcsname{%
#1{\mathpalette{\sla@{#2}{#3}{#4}}{#5}}}}
 \catcode`\@=12
\declareslashed{}{/}{.08}{0}{D}
 \declareslashed{}{/}{.1}{0}{A}
 \declareslashed{}{/}{0}{-.05}{k}
 \declareslashed{}{/}{.1}{0}{\partial}
 \declareslashed{}{\not}{-.6}{0}{f}

\def\lsim{\mathrel {\vcenter {\baselineskip 0pt \kern 0pt
    \hbox{$<$} \kern 0pt \hbox{$\sim$} }}}
\def\gsim{\mathrel {\vcenter {\baselineskip 0pt \kern 0pt
    \hbox{$>$} \kern 0pt \hbox{$\sim$} }}}

\def\zp{Z^\prime}
\newcommand{\bea}{\begin{eqnarray}}
\newcommand{\eea}{\end{eqnarray}}

\begin{document}

\baselineskip=15pt
\preprint{}

\title{Tau lepton charge asymmetry and new physics at the LHC}

\author{Sudhir Kumar Gupta and German Valencia}

\email{skgupta@iastate.edu,valencia@iastate.edu}

\affiliation{Department of Physics, Iowa State University, Ames, IA 50011.}

\date{\today}

\vskip 1cm
\begin{abstract}

We consider the possibility of studying new physics that singles out the tau-lepton at the LHC. We concentrate on the tau-lepton charge asymmetry in  $\tau^+\tau^-$ pair production as a tool to probe this physics beyond the Standard Model. We  consider two generic scenarios for the new physics. We first study a non-universal $Z^\prime$ boson as an example of a new resonance that can single out tau-leptons.  We then consider vector lepto-quarks coupling of the first generation quarks with the third generation leptons  as an example of non-resonant new physics. We find that in both cases the charge asymmetry can be sufficiently sensitive to the new physics to provide useful constraints at the LHC. 

\end{abstract}

\pacs{PACS numbers: }

\maketitle

\section{Introduction}

The Large Hadron Collider (LHC) has been running since earlier last year in an initial phase of rediscovering the Standard Model (SM).  A variety of observables are necessary to completely measure the SM couplings and we concentrate here on the charge asymmetry, the hadron collider equivalent of the familiar forward-backward asymmetry. During the early stages of LHC running, the available event rates will possibly allow  measurements of lepton charge asymmetries near the $Z$ peak. These translate into measurements of the weak angle, $\sin^2\theta_W$, and can help in the SM rediscovery phase at LHC. 

When the LHC is running at its design energy and luminosity, it will be possible to extend these measurements to higher regions of dilepton invariant mass, $M_{\ell \ell}$, not probed by LEP II, and in this way play an important role in the search for new physics. The Drell-Yan dilepton pair production process is important in this context due to its clean signature.  
In this paper we discuss the use of the lepton charge asymmetry as a
tool to discover and identify new physics, with emphasis on the $\tau$-pair production channel.

There exists a vast literature dedicated to the study of new physics associated with the top-quark, that is motivated by its large mass and possible unique role in electroweak symmetry breaking. It is natural to ask whether new physics that singles out the top-quark does not in fact single out the whole third generation of the SM fermions. This is particularly the case in light of the existing anomaly in the forward-backward asymmetry of the $b$-quark \cite{afb-b}. 

With this in mind, we wish to explore the possibility of new physics that affects the tau-lepton but does not show up in studies of muons or electrons. To this effect we consider two different scenarios that single out the third generation leptons. The scenarios are not complete models, but instead they describe two simple possibilities. Our first example is a non-universal $Z^\prime$ which has been studied before in connection with the top-quark \cite{Han:2004zh}.  Here we explore its consequences in tau-lepton physics at the LHC in the large tau-pair invariant mass, $M_{\tau\tau}$ region. It is possible that such a $Z^\prime$ can be detected by simply looking for bumps in the $M_{\tau\tau}$ distribution, but we emphasize here the question of detecting its effect via the lepton charge asymmetry.  Our second example consists of vector lepto-quarks which provide a benchmark for effects from non-resonant new physics. By associating the third family of leptons with the first family of quarks, the lepto-quarks in question single out the tau-lepton \cite{Valencia:1994cj}.

We first present a study at the $\tau$-lepton level in Section III. In Section IV, we back up our conclusions by considering selected $\tau$-lepton decay modes.

\section{Charge Asymmetry (Forward-Backward asymmetry)}

Forward-backward asymmetries have proved to be valuable tools for constraining the SM and searching for new physics  in $e^+e^-$ colliders. A prominent example being the $A_{FB}^b$  anomaly measured at LEP  \cite{:2005ema}, which remains a hint for new physics \cite{afb-b}. More recently the Tevatron has reported an anomaly in the forward-backward asymmetry of the top-quark \cite{afb-top} which has also received considerable attention \cite{afb-top-th}.

Forward-backward asymmetries can still be defined for the LHC even though it is a symmetric $pp$ collider. The idea is to consider processes that are initiated by $q\bar{q}$ annihilation at the parton level. It is then possible to define the forward-backward asymmetry in the usual way in the parton center-of-mass frame (CM): the forward direction corresponding to the incoming quark direction.  To connect with the $pp$ collider, it is sufficient to recall that the quarks in the proton carry, on average, a larger fraction of the proton momentum than the anti-quarks, the direction of the quark momentum is thus correlated with the direction of the total momentum of the event in the lab frame (the boost direction). 

Specifically, one can start from the parton CM asymmetry defined in the usual way,
\begin{eqnarray}
{\cal A}_{FB}^\star(q\bar{q} \to \ell^- \ell^+)& \equiv & \frac{\sigma_F - \sigma_B}{\sigma_F + \sigma_B} 
\label{afbstar}
\end{eqnarray}
where $\sigma_F$ ($\sigma_B$) is the respective cross-section for producing a lepton $\ell^-$ that travels forward (backward) with respect to the initial quark direction. This has been studied by both the ATLAS \cite{Aharrouche:2007mq} and CMS \cite{cmsnote} collaborations in connection with measurements of $\sin^2\theta_W$ from the SM asymmetries at the $Z$ peak.

In the lab frame, a non-vanishing ${\cal A}^\star_{FB}$ will manifest itself as a rapidity asymmetry: the forward leptons in the parton center-of-mass will have a larger rapidity in the lab frame. At the LHC, the symmetry of the initial $pp$ state results in a quark direction that is equally likely to correspond to either proton. The net effect of an ${\cal A}^\star_{FB}$ in the lab frame at the LHC is a charge asymmetry in which the type of lepton that preferred the backward direction in the CM now concentrates in the central rapidity region. This charge asymmetry is in a sense equivalent to a forward backward asymmetry, but defined in lab frame for a symmetric $pp$ collider. \footnote{This charge asymmetry should not be confused with the unequal production rates for $W^+$ and $W^-$ in $pp$ colliders.}

Following this argument, it is common to define a charge asymmetry in terms of the fermion rapidity \cite{Ferrario:2009ns,Catani:2010en}. In our case,
\begin{equation}
{\cal A}(y) = \frac{N_{\ell^+}(y) - N_{\ell^-}(y)}{N_{\ell^+}(y) + N_{\ell^-}(y)}
\label{eq:cha}
\end{equation}
where $y$ is rapidity of the lepton $\ell^\pm$ and $N$ is the number of events with a given rapidity $y$. One advantage of using this charge asymmetry over the forward-backward asymmetry in Eq.~\ref{afbstar} is that the parton CM frame does not need to be reconstructed, as this reconstruction is not always possible. It is also common to define an integrated asymmetry over a central region, limited by $y_c$:
\begin{equation}
{\cal A}_c (y_c) = \frac{N_{\ell^+}(- y_c \leq y \leq y_c) -  N_{\ell^-}(- y_c \leq y \leq y_c)} 
                                       {N_{\ell^+}(- y_c \leq y \leq y_c) + N_{\ell^-}(- y_c \leq y \leq y_c)}.                                      
\label{ycasym}
\end{equation}
This integrated asymmetry can be optimized with a carefully chosen $y_c$ \cite{Ferrario:2009ns}.  Notice that the symmetry of the initial $pp$ state at LHC causes the integrated asymmetry to vanish when the whole rapidity range is used. Finally, it may be convenient to integrate the charge asymmetry over different ranges of $M_{\ell\ell}$ as we discuss below.

\section{Model Descriptions and Numerical Analysis}

In this section we  describe briefly the two models we use to illustrate the effectiveness of the charge asymmetry and we present the corresponding numerical results. These two models are chosen because we are mostly interested in applications to $\tau$-lepton physics at LHC and both are examples of new physics that singles out the $\tau$-lepton.  

In all cases we use  {\tt MadGraph 4.4.39} \cite{madgraph} for event generation and {\tt PYTHIA 6.4.21} \cite{pythia} for the analysis. To this end we implement the vertices that originate in each of the two new physics models directly into  {\tt MadGraph}. We use {\tt CTEQ6L-1} parton distribution functions~\cite{Pumplin:2002vw}.  
We also implement the basic acceptance cuts $p_{T_\tau} > 20$ GeV, $\left|\eta_\tau \right| < 2.5$, and $\Delta R_{\tau\tau} > 0.4$. Our results will show that the measurements we propose will not be possible in the early running of LHC. 
For this reason  we will assume a $\tau$ physics program when LHC is running at $\sqrt{S} = 14$~TeV  center-of-mass energy, and with an integrated luminosity of $10~{\rm fb}^{-1}$ per year. 

We first carry out our analysis at the $\tau$-lepton level, without concerning ourselves with the subsequent $\tau$ decays. This will serve to assess the best case scenario of high efficiency $\tau$-lepton tagging and reconstruction as both CMS and ATLAS expect to be able to detect $\tau$-leptons with relatively large efficiencies \cite{CMS,Atlas}. We will then partially address these issues by considering a few selected $\tau$-lepton decay modes.

The SM itself produces a non-zero charge asymmetry so that a search for new physics involves measuring the deviation from the SM.  Depending on the new physics, this deviation may be small and a precise measurement may be needed. 

\begin{figure}%[htb]
\centerline{  
\includegraphics[angle=-90, width=.6\textwidth]{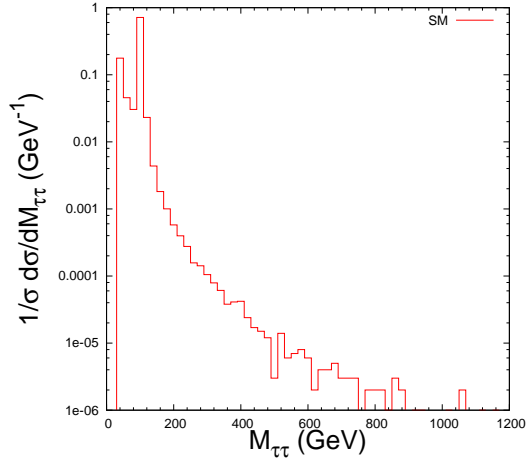}
}
\caption{\small\sf Dilepton invariant mass distribution for the process $pp\to \tau^+ \tau^-$ within standard model at the LHC for $\sqrt{S} = 14$ TeV.}
\label{fig:mllsm}
\end{figure}

We begin with a discussion of the charge asymmetry in the SM. Within the SM, dileptons at the LHC are  produced predominantly via $s$-channel exchange of a photon or a $Z$ boson with a total cross-section (for our acceptance cuts) of $\sim 941$~pb at $\sqrt{S} = 14$~TeV. The corresponding SM $\tau^+\tau^-$ events exhibit an $M_{\tau\tau}$ distribution shown in Figure~\ref{fig:mllsm}. This differential cross-section exhibits a clear $Z$ peak and falls rapidly with $M_{\tau\tau}$. To distinguish possible new physics it is therefore useful to exclude the $Z$ region and to look as far out in $M_{\tau\tau}$ as the event rate permits. We will find that the best region to look for new physics (at least in the two examples we consider) is $M_{\tau\tau} > 200$~GeV.

\begin{figure}%[htb]
\centerline{  
\includegraphics[angle=-90, width=.6\textwidth]{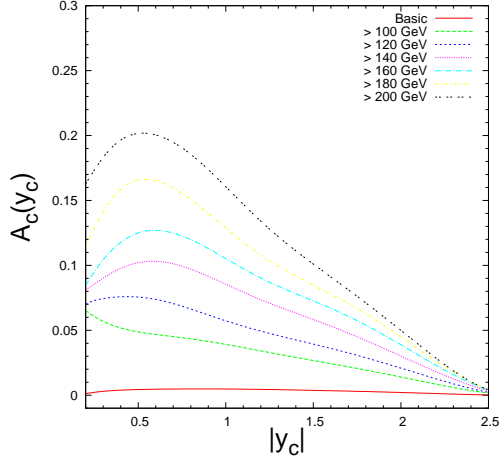}
}
\caption{\small\sf Integrated charge asymmetry for the standard model with basic acceptance cuts as well as different  minimum dilepton invariant mass cuts.}
\label{fig:acsm}
\end{figure}

Our main observable is the integrated charge asymmetry over both rapidity, within a range determined by $|y| < y_c$, and $M_{\tau\tau} > M_{min}$. Previous studies related to the top-quark have shown that there is an optimal value for $y_c$ in Eq.~\ref{ycasym}~\cite{Ferrario:2009ns} and we illustrate this point within the context of the SM in Figure~\ref{fig:acsm}. This figure indicates that values around $y_c \sim 0.5$ maximize the integrated charge asymmetry more or less independently from the $M_{min}$ value used. The selection of $M_{min}$ proceeds as discussed above: we want to exclude the $Z$ peak region to minimize the effect of the SM as a background to the new physics without reducing the event rate too much. It turns out that the asymmetry increases with increasing $M_{min}$ (at least up to the $M_{min}=300$~GeV that we have tested) and this is illustrated in Figure~\ref{fig:acsm}. As $M_{min}$  increases, this cut is more effective in rejecting events with a lower boost and this results in an increased asymmetry. The choice of a specific value for $M_{min}$ will thus result from a compromise between a somewhat larger asymmetry and reduced statistics as $M_{min}$ increases. We explore this further  when we discuss results within the illustrative models.

%--------------------------------------------------------------------
%--------------------------------------------------------------------
%--------------------------------------------------------------------
%--------------------------------------------------------------------

\subsection{Non-Universal $Z^\prime$ model}

One of the most frequently studied extensions of the SM is an additional $U(1)^\prime$ symmetry and its associated $Z^\prime$ boson \cite{review}. Although the $U(1)^\prime$ charges are family universal in most of the models discussed in the literature, this need not be the case. It is well known that a non-universal $Z^\prime$ induces tree-level flavor changing neutral currents (FCNC) which are severely constrained by experiment, most notably meson mixing  \cite{fcncconstraints}.

In this paper we will not concern ourselves with the FCNC, but rather with the possibility of an enhanced coupling to the $\tau$-lepton as in the models of Ref.~\cite{b3ltau,heval}, that can be probed at LHC. 

We write the general couplings of a non-universal $Z^\prime$ boson to the SM fermions as follows,
\begin{eqnarray}
{\cal L}_{\zp} &=&
 {g\over 2\cos\theta_W} \,
\left(\bar{f}\gamma^\mu \left(c^f_L P_L + c^f_R P_R \right) f \right) Z^{\prime_\mu}.\, 
\label{zplag}
\end{eqnarray}
A model with an enhanced coupling $c^\tau_{L,R}$ singles out the $\tau$-lepton pair production process. Of course, at LHC, we also need to know how the $Z^\prime$ is produced and this forces us to specify its couplings to light quarks.

We consider enhanced $Z^\prime$ couplings to the third generation that are accompanied by correspondingly suppressed couplings to the light generations as happens in the models mentioned above. This results in a complete $q\bar{q}\to \tau^+\tau^-$ process that is of electroweak strength. Existing constraints on this $Z^\prime$ are twofold. From the LEP-II process $e^+e^- \to \tau^+\tau^-$ considering both the cross-section and the forward-backward asymmetry as a function of $M_{\tau\tau}$, Ref.~\cite{heval} concludes that masses $M_{Z^\prime} \lsim 500$~GeV are excluded.  The phenomenology of such a $Z^\prime$ at LHC vis-a-vis its enhanced couplings to the top-quark has also been considered both for flavor conserving \cite{Han:2004zh}, as well as flavor violating processes \cite{FCNCatLHC}. It was found that the simple Drell-Yan-type processes are completely overwhelmed by QCD background. Processes with three or four top (or bottom) quarks at the LHC could constrain the model for $Z^\prime$ masses up to 2 TeV, but several hundred fb$^{-1}$ of integrated luminosity would be necessary. 

\begin{figure}%[htb]
\centerline{
\includegraphics[angle=0, width=.4\textwidth]{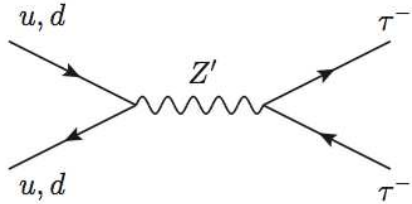}
}
\caption{\small\sf Representative Feynman diagram for dilepton production through $Z^\prime$ exchange.}
\label{fig:fgzp}
\end{figure}

In this model, the additional contribution to the partonic process $q{\bar q}\to \ell^+\ell^-$ 
is due to an s-channel $Z^{\prime}$ exchange as shown in the Figure~\ref{fig:fgzp}. 
We consider two cases, corresponding to $M_{Z^\prime}= 600$~GeV and $M_{Z^\prime}= 1$~TeV. In both cases we will use a relatively narrow $Z^\prime$ width of 100~GeV. For illustration purposes we use purely right handed couplings  with overall strength $c^u_R\cdot c^\tau_R = 1/3$.

%--------------------------------------------------------------------
%--------------------------------------------------------------------

%--------------------------------------------------------------------
\begin{table}[th] 
\begin{center} 
\begin{tabular}{|c|r@{ $\pm$ }l|r@{$\pm$}l|r@{$\pm$}l|} 
\hline\hline 
\multirow{2}{*}{Cuts } & \multicolumn{4}{|c|}{$\zp$ Model} & \multicolumn{2}{|c|}{SM} 
\\\cline{2-5}
& \multicolumn{2}{|c|}{$M_{\zp} = 0.6$} & \multicolumn{2}{|c|}{$M_{\zp}= 1$} &\multicolumn{2}{|c|}{} \\ 
\hline \hline
Basic &948.9&1.3&948.6&1.3&941.0&1.2\\\hline
$M_{\tau\tau} > 0.2$ TeV &1.58&0.002&1.49&0.002&1.41&0.002\\
\hline \hline
\end{tabular} 
\end{center}
\caption{\small\sf Dilepton pair cross-sections (in pb) for  the $\zp$ model and the SM at the LHC with $\sqrt{S} = 14$ TeV. All the masses shown here are given in TeV.}
\label{t:sigmazp}
\end{table} 
%--------------------------------------------------------------------

\begin{figure}%[htb]
\centerline{  
\includegraphics[angle=-90, width=.6\textwidth]{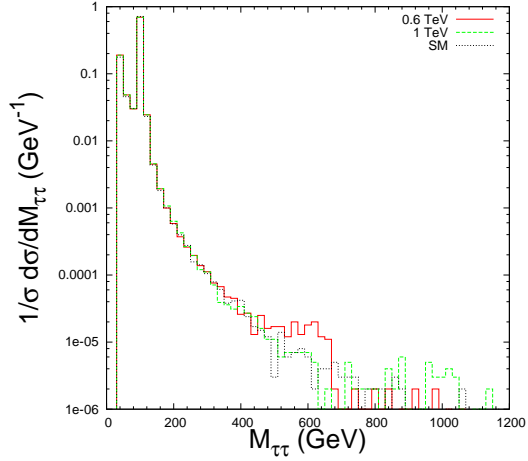}
}
\caption{\small\sf Dilepton invariant mass distributions for the 
$\zp$-model at the LHC with $\sqrt{S} = 14$ TeV for $M_{\zp} = 0.6$ and 1 TeV.}
\label{fig:mllzp}
\end{figure}

In Figure~\ref{fig:mllzp} we present the resulting $M_{\tau\tau}$ distribution for the two values of the $Z^\prime$ mass, 600~GeV and 1~TeV, as well as for the SM.  As Table~\ref{t:sigmazp} shows, removing the low $M_{\tau\tau}$ region results in a significant loss of events. However, Figure~\ref{fig:mllzp} illustrates that in the region of low ditau invariant mass the SM completely obscures the new physics. To increase the sensitivity to new physics we therefore need to remove this low invariant ditau mass region, especially the $Z$ peak. 

\begin{figure}[htb]
\centerline{
\includegraphics[angle=-90, width=.6\textwidth]{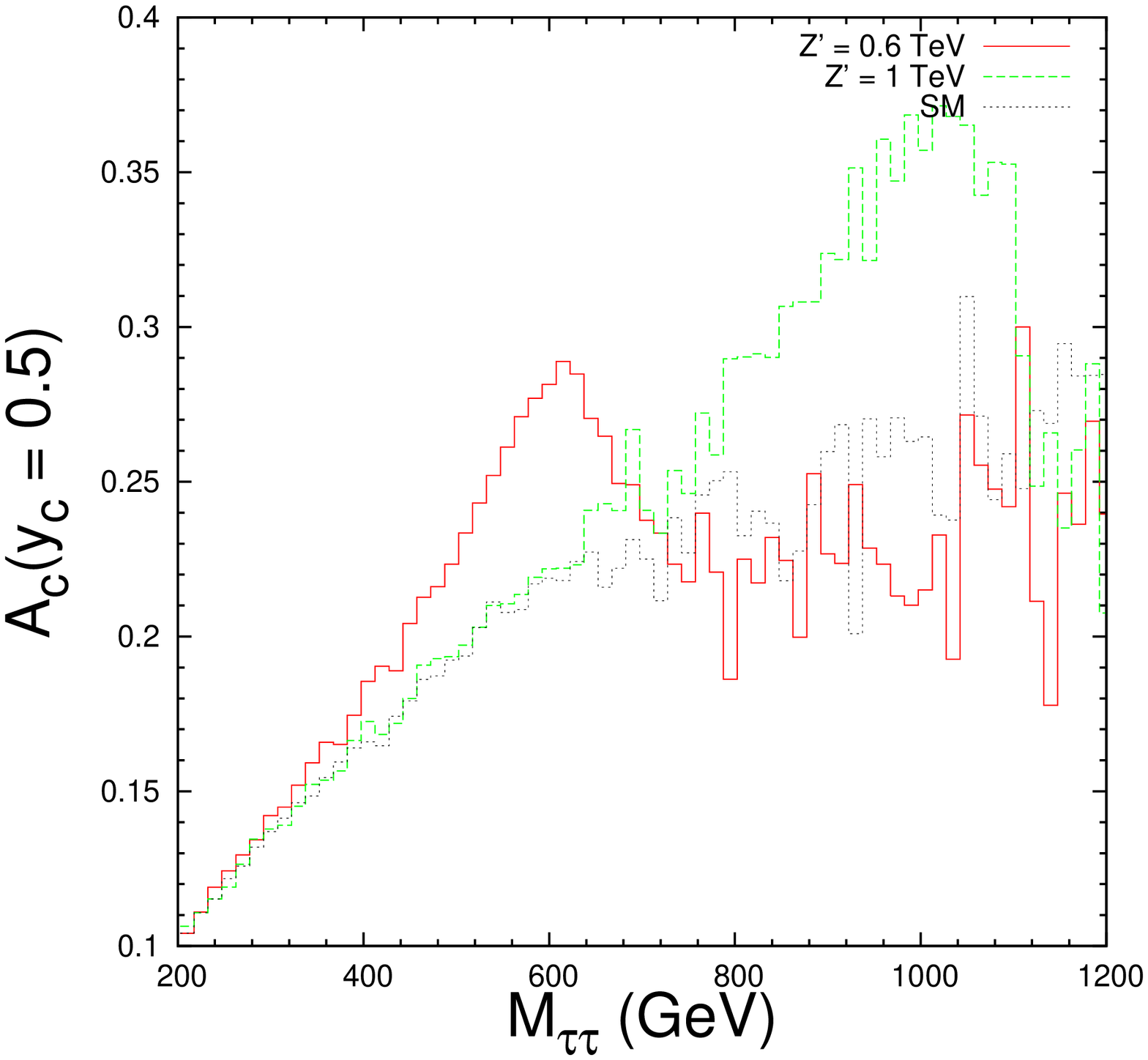}
\includegraphics[angle=-90, width=.6\textwidth]{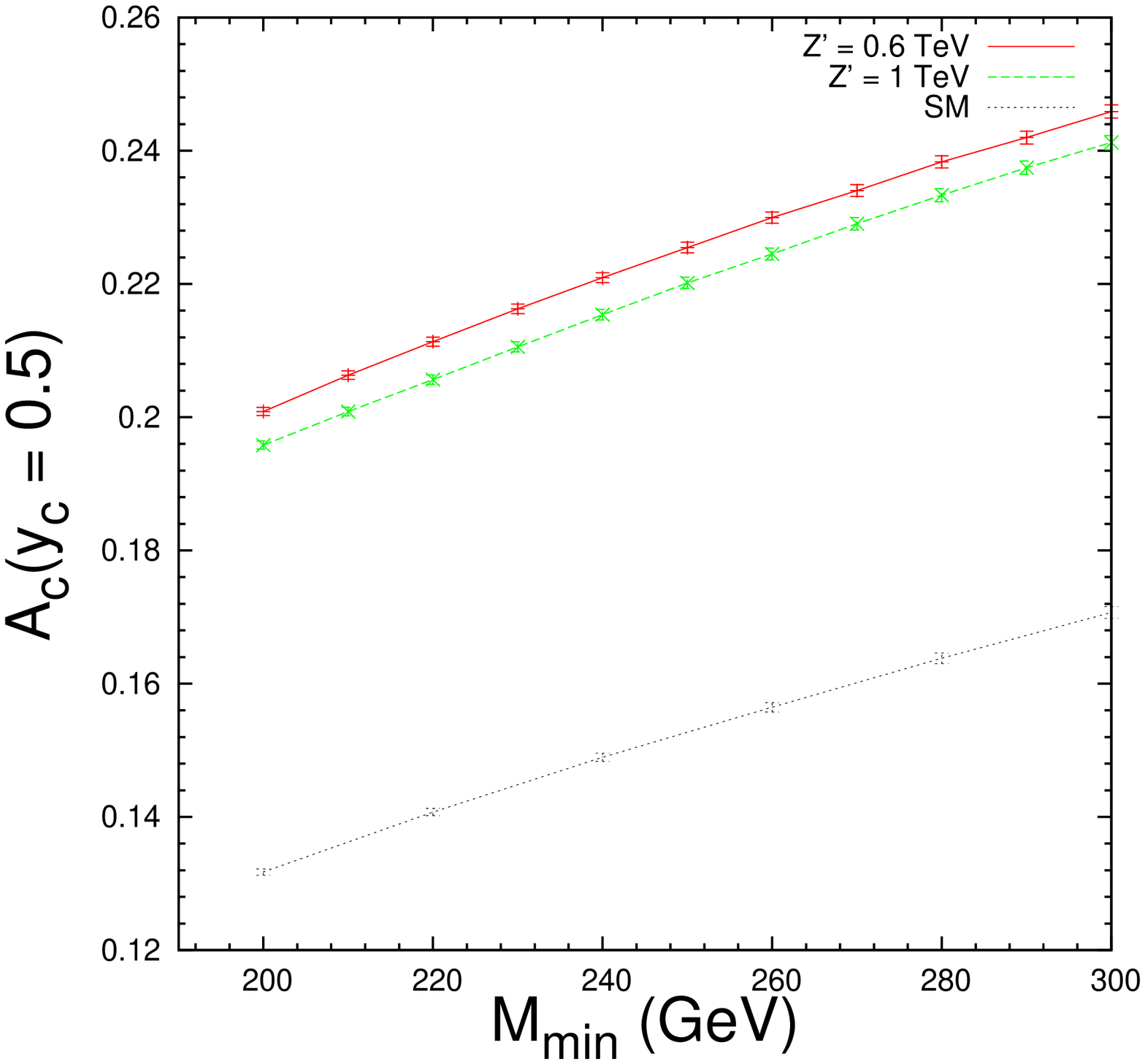}
}
\caption{\small\sf Lepton charge asymmetry in the $Z^\prime$ model for $M_{\tau\tau} > 200$~GeV, $y_c = 0.5$ and $M_{Z^\prime} = 0.6$ and 1 TeV vs $M_{\tau\tau}$ (left) and integrated over $M_{\tau\tau} \ge M_{min}$ (right). This model has only non-zero couplings as $c^{\tau}_R$, and $c^u_R$ such that $c^{\tau}_R \cdot c^u_R = 1/3$.}
\label{fig:mizp}
\end{figure}

If there is a new resonance, such as a $Z^\prime$, with a mass in the energy range accessible to LHC it could be discovered by studying the cross-section or an invariant mass distribution such as that shown in Figure~\ref{fig:mllzp}. We do not address this possibility in this paper as this is not the focus of our study. Our focus is the charge asymmetry: if sufficiently different from the SM prediction it could signal new physics that is not directly observable as a resonance (for example, because it is too heavy). The charge asymmetry is a very useful observable even if new physics is first discovered as a resonance: it would serve to distinguish between different possibilities as we illustrate below. We therefore begin by asking whether the $Z^\prime$ can yield a charge asymmetry that differs sufficiently from the SM to be observed. This is illustrated in Figure~\ref{fig:mizp}.
%--------------------------------------------------------------------
%--------------------------------------------------------------------

%--------------------------------------------------------------------
%--------------------------------------------------------------------

The first figure shows the charge asymmetry for $y < y_c = 0.5$ as a function of dilepton invariant mass. The second figure shows the corresponding asymmetry  integrated over dilepton invariant mass, as a function of $M_{min}\gsim 200$~GeV.

\begin{figure}[htb]
\centerline{
\includegraphics[angle=-90, width=.6\textwidth]{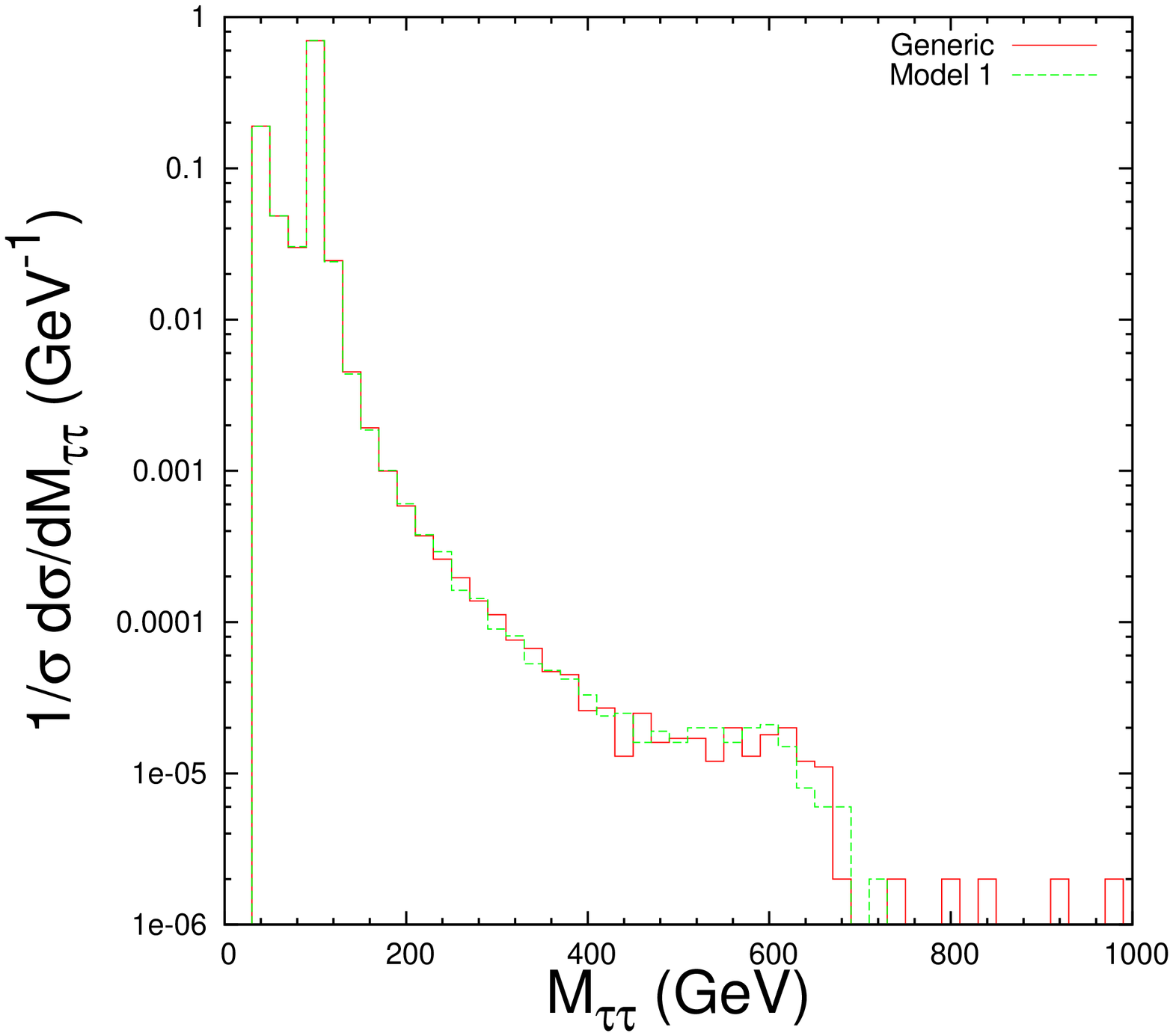}
\includegraphics[angle=-90, width=.6\textwidth]{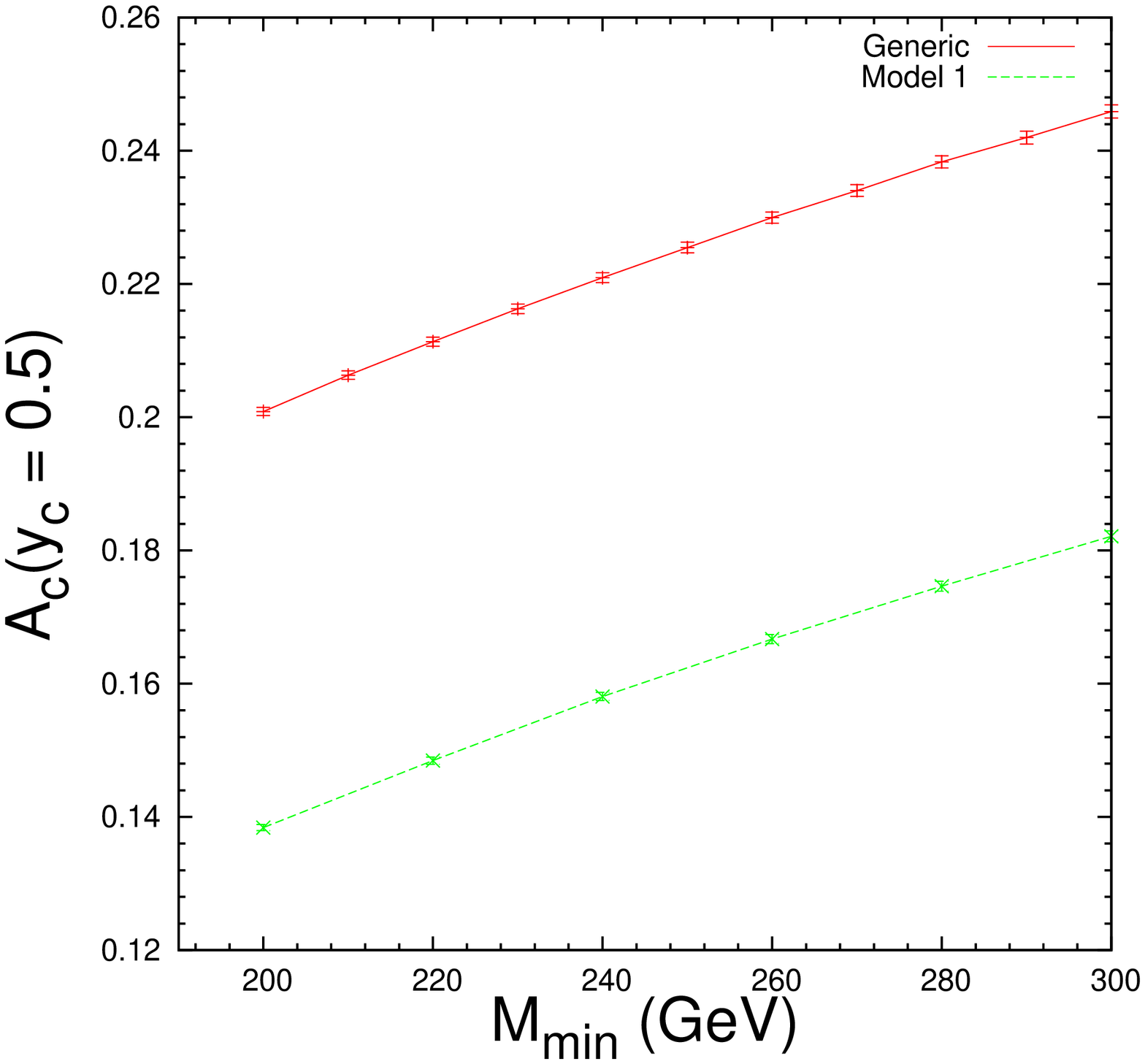}
}
\caption{\small\sf Comparison of two $Z^\prime$ models with $M_{Z^\prime}=600$~GeV in invariant mass distribution $M_{\tau\tau}$ (left) and integrated charge asymmetry (right). The model labeled `Generic'  has only non-zero couplings $c^{\tau}_R$, and $c^u_R$ such that $c^{\tau}_R \cdot c^u_R = 1/3$. The model labeled `Model 1'  is that of Ref.~\cite{Han:2004zh}.}
\label{fig:Zpcompare}
\end{figure}

We have explored the effect of variations in the couplings to some extent. Coupling the $Z^\prime$ to $d$-quarks instead of $u$-quarks for its  production process diminishes its contributions at LHC. Replacing the purely right-handed coupling with a purely left-handed coupling affects the interference between the SM and the $Z^\prime$.  In Figure~\ref{fig:Zpcompare} we emphasize  the utility of the charge  asymmetry in helping to untangle any possible new physics that may first be observed in the invariant mass distribution. We compare two $Z^\prime$ cases both with mass 600~GeV: the generic case described above, where the only non-zero couplings are $c^{\tau}_R$, and $c^u_R$ such that $c^{\tau}_R \cdot c^u_R = 1/3$; and the model studied in Ref.~\cite{Han:2004zh}. As is evident from the figure, the two models have nearly identical invariant mass distributions but significantly different integrated charge asymmetry.

%--------------------------------------------------------------------
%--------------------------------------------------------------------

To quantify these results further we study the statistical sensitivity of the LHC to these signals. In Table~\ref{t:acyzp} we show the integrated $\tau$-lepton charge asymmetry in percent for the two values of the $Z^\prime$  mass as well as for the SM. We show results for several values of $M_{min}$ for $y_c=0.5$ with their corresponding 1$\sigma$ statistical error. These errors correspond to accumulated statistics for 10 fb$^{-1}$ at $\sqrt{S}=14$~TeV. We can see that the asymmetry increases (both for the SM and the $Z^\prime$ model) as we increase  $M_{min}$, but the error also increases due to the correspondingly reduced statistics. 

%--------------------------------------------------------------------
\begin{table}[th] 
\begin{center} 
\begin{tabular}{|c|r@{ $\pm$ }l|r@{$\pm$}l|r@{$\pm$}l|} 
\hline\hline 
\multirow{2}{*}{Cuts } & \multicolumn{4}{|c|}{$\zp$ Model} & \multicolumn{2}{|c|}{SM} 
\\\cline{2-5}
& \multicolumn{2}{|c|}{$M_{\zp} = 0.6$} & \multicolumn{2}{|c|}{$M_{\zp}= 1$} &\multicolumn{2}{|c|}{} \\ 
\hline \hline
$M_{\tau\tau} > 0.20$ TeV& 20.1 & 0.6 & 19.6 & 0.6 & 13.2 & 0.6\\
$M_{\tau\tau} > 0.22$ TeV& 21.1 & 0.7 & 20.6 & 0.7 & 14.1 & 0.7\\
$M_{\tau\tau} > 0.24$ TeV& 22.1 & 0.8 & 21.5 & 0.8 & 14.9 & 0.8\\
$M_{\tau\tau} > 0.26$ TeV& 23.0 & 0.8 & 22.5 & 0.9 & 15.6 & 0.9\\
$M_{\tau\tau} > 0.28$ TeV& 23.8 & 1.0 & 23.3 & 1.0 & 16.4 & 1.0\\
$M_{\tau\tau} > 0.30$ TeV& 24.6 & 1.0 & 24.1 & 1.1 & 17.1 & 1.1\\
\hline \hline
\end{tabular} 
\end{center}
\caption{\small\sf Integrated lepton charge asymmetry (in percent) , ${\cal A}_c (y_c)$, for the $\zp$ model and the SM. The $1 \sigma$-errors correspond to statistics for one year of LHC data (at $\int {\cal L} dt = 10$ fb$^{-1}$ per year). All the masses here are given in TeV. In this Table the $Z^\prime$ model's only non-zero couplings are $c^{\tau}_R$, and $c^u_R$ such that $c^{\tau}_R \cdot c^u_R = 1/3$.}
\label{t:acyzp}
\end{table} 
%--------------------------------------------------------------------

\subsection{Lepto-quark Models}

As a second example of new physics that singles out $\tau$-lepton pairs at LHC we consider lepto-quarks. This example will serve to illustrate the case where the new physics does not have a resonant peak in the channel of interest.

Generic couplings of vector lepto-quarks to standard model fermions can be written in the form \cite{generalLQ}
\begin{eqnarray}
{\cal L}_{LQ} &=& {\cal L}_{SM} \\\nonumber
&+& \lambda^{(R)}_{V_0}\cdot \overline{d} \gamma^{\mu} P_R e
\cdot V_{0\mu}^{R\dagger} +
%%%%%%%%%%%%%%%%%%%%%%%%%%%%%%%%%%%
\lambda^{(R)}_{\tilde V_0}\cdot \overline{u} \gamma^{\mu} P_R e \cdot \tilde{V}_{0\mu}^{\dagger} \\ \nonumber
%%%%%%%%%%%%%%%%%%%%%%%%%%%%%%%%%%%
&+& \lambda^{(R)}_{V_{1/2}}\cdot \overline{d^c} \gamma^{\mu} P_L \ell
\cdot {V}_{1/2\mu}^{R\dagger} +
%%%%%%%%%%%%%%%%%%%%%%%%%%%%%%%%%%%
\lambda^{(R)}_{\tilde V_{1/2}}\cdot \overline{u^c} \gamma^{\mu} P_L \ell \cdot
\tilde{V}_{1/2\mu}^{\dagger} \\ \nonumber
%%%%%%%%%%%%%%%%%%%%%%%%%%%%%%%%%%%
&+&
\lambda^{(L)}_{V_0}\cdot \overline{q} \gamma^{\mu} P_L \ell \cdot
V_{0\mu}^{L\dagger} +
%%%%%%%%%%%%%%%%%%%%%%%%%%%%%%%%%%%
\lambda^{(L)}_{V_{1/2}}\cdot \overline{q^c} \gamma^{\mu} P_R e \cdot
V_{1/2\mu}^{L\dagger} \\ \nonumber
%%%%%%%%%%%%%%%%%%%%%%%%%%%%%%%%%%%
&+& \lambda^{(L)}_{V_1}\cdot \overline{q} \gamma^{\mu}  P_L
{V}_{1\mu}^{\dagger} \ell + h.c..
\label{eq:laglq}
\end{eqnarray}

In the equation above $V_i^{j}$ are vector lepto-quarks with weak isospins $i=0, 1/2, 1$ 
coupled to left-handed ($j = L$) or right-handed ($j = R$) quarks respectively. The cases with weak isospin $i=0,1/2$ admit two possible values of hypercharge (leading to different electric charges), we denote the two possibilities by $V$ and $\tilde{V}$. Scalar lepto-quarks are also possible \cite{scalarLQ}.

For our study, we have in mind Pati-Salam \cite{Pati:1974yy} type lepto-quarks in which the coupling $\lambda^{(R)}_{V_0}=\lambda^{(L)}_{V_0} = g_s/\sqrt{2}$ at the unification scale. In order to single out the $\tau$-lepton we envision a scenario described in Ref.~\cite{Valencia:1994cj} in which the third generation of leptons is associated with the first generation of quarks.  

For our numerical analysis we will only consider the following two types of lepto-quark as an illustration:
\begin{itemize}

\item LQ-1, a Pati-Salam lepto-quark as described above with coupling $\lambda^{(R)}_{V_0}=\lambda^{(L)}_{V_0} = g_s/\sqrt{2}$ between the first generation quarks and the third generation leptons. This leads to $\tau^+\tau^-$ pairs produced by $\bar{d}d$ annihilation at LHC.

\item LQ-2, with coupling $\lambda^{(R)}_{\tilde{V}_0} = g_s/\sqrt{2}$ between the first generation quarks and the third generation leptons. This one is a variation of LQ-1 in which the $pp\to \tau^+\tau^-$ process is initiated by $u\bar{u}$ annihilation (instead of $d\bar{d}$ in LQ-1).

\end{itemize}

Although lepto-quarks with these particular flavor quantum numbers are not commonly discussed, generic direct bounds on vector lepto-quark masses are in the hundreds of GeV range \cite{pdb}. It is well known, however that rare decays place stronger indirect constraints, for example \cite{lepto-quarkconst}. For the specific flavor couplings we use here, Ref.~\cite{Valencia:1994cj} identified the process $B_s\to \mu e$  as the one yielding the tightest constraint. The most recent number obtained using this process corresponds to a lower bound on the lepto-quark mass near 50~TeV \cite{CDFLQ}. Although one must keep these indirect constraints in mind, they are not fool-proof substitutes for direct searches; instead they yield complementary information.

In both scenarios we will show numerical results for lepto-quark masses of 600~GeV and 1~TeV which are within the direct reach of LHC but significantly below the indirect limits. 

\begin{figure}%[htb]
\centerline{
\includegraphics[angle=0, width=.6\textwidth]{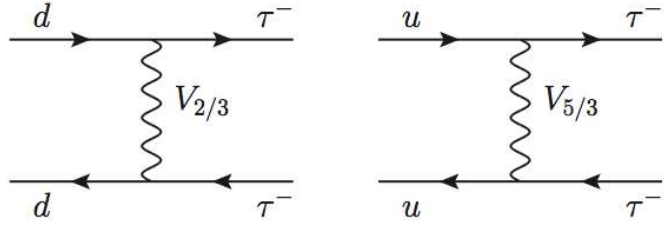}
}
\caption{\small\sf Feynman diagram for dilepton production through lepto-quark exchange in lepto-quark Model 1 (Left) and 2 (Right).}
\label{fig:fglq}
\end{figure}

The additional contributions to the dilepton cross-section are due to t-and u-channel exchange of lepto-quarks as illustrated in the Feynman  diagrams shown in Fig.~\ref{fig:fglq}. In Table~\ref{t:sigmalq} we compare the $\tau$-pair cross-section that results in these models with the SM.

%--------------------------------------------------------------------
%--------------------------------------------------------------------

%--------------------------------------------------------------------
\begin{table}[th] 
\begin{center} 
\begin{tabular}{|c|r@{ $\pm$ }l|r@{$\pm$}l|r@{$\pm$}l|r@{$\pm$}l|r@{$\pm$}l|} 
\hline\hline 
\multirow{2}{*}{Cuts } & \multicolumn{4}{|c|}{LQ-1} & \multicolumn{4}{|c|}{LQ-2 } & \multicolumn{2}{|c|}{SM} 
\\\cline{2-9}
& \multicolumn{2}{|c|}{$M_{V_{2/3}} = 0.6$} & \multicolumn{2}{|c|}{$M_{V_{2/3}}= 1$} & 
\multicolumn{2}{|c|}{$M_{V_{5/3}} = 0.6$} & \multicolumn{2}{|c|}{$M_{V_{5/3}} = 1$} & 
\multicolumn{2}{|c|}{} \\ 
\hline \hline
Basic&938.1&1.3&938.4&1.3&918.5&1.4&912.9&1.3&941.0&1.2\\\hline
$M_{\tau\tau} > 0.2$ TeV &2.06&0.002&1.43&0.002&4.70&0.002&2.21&0.002&1.41&0.002\\
\hline \hline
\end{tabular} 
\end{center}
\caption{\small\sf Dilepton pair cross-sections (in pb) for lepto-quark model 1, 2 (LQ-1, 2) and, the SM at the LHC with $\sqrt{S} = 14$ TeV. All the masses shown here are given in TeV.}
\label{t:sigmalq}
\end{table} 
%--------------------------------------------------------------------

%--------------------------------------------------------------------
\begin{figure}%[htb]
\centerline{  
\includegraphics[angle=-90, width=.6\textwidth]{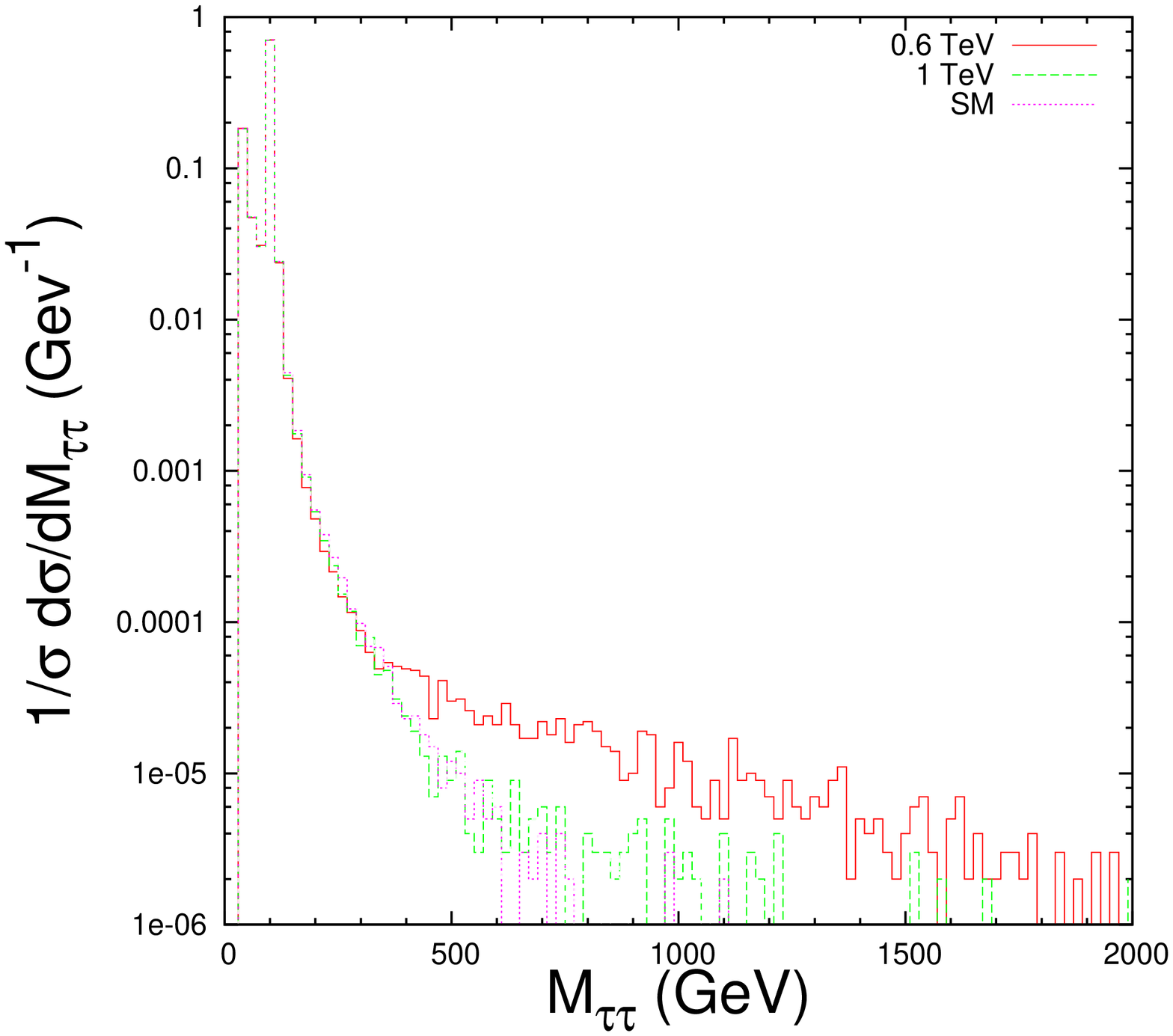}
\hskip -2.2cm
\includegraphics[angle=-90, width=.6\textwidth]{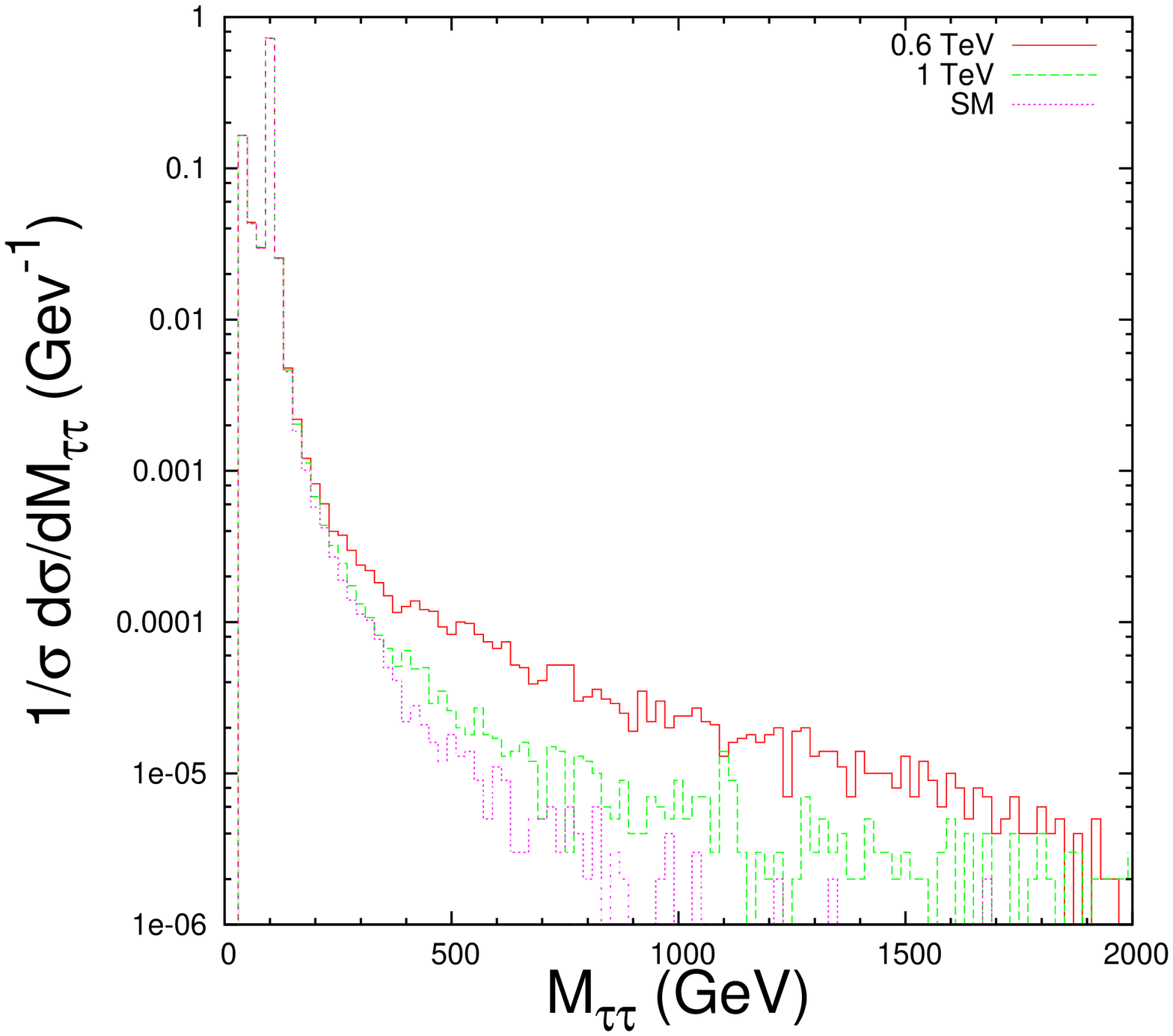}
}
\caption{\small\sf Dilepton invariant mass distributions for the scenarios LQ-1 and LQ-2 at the LHC with $\sqrt{S} = 14$ TeV for $M_{LQ} = 0.6$ and 1 TeV.}
\label{fig:mlllq1}
\end{figure}
%--------------------------------------------------------------------

In Figure~\ref{fig:mlllq1} we show the dilepton invariant mass distributions that result in both cases LQ-1 and LQ-2 for $M_{LQ} = 0.6, 1$~TeV respectively. As expected, these distributions do not show any ``bumps'' as the lepto-quarks are not exchanged in the s-channel. However, they do show an enhancement over the SM, particularly for the larger values of $M_{\tau\tau}$.

\begin{figure}%[htb]
\centerline{
\includegraphics[angle=-90, width=.6\textwidth]{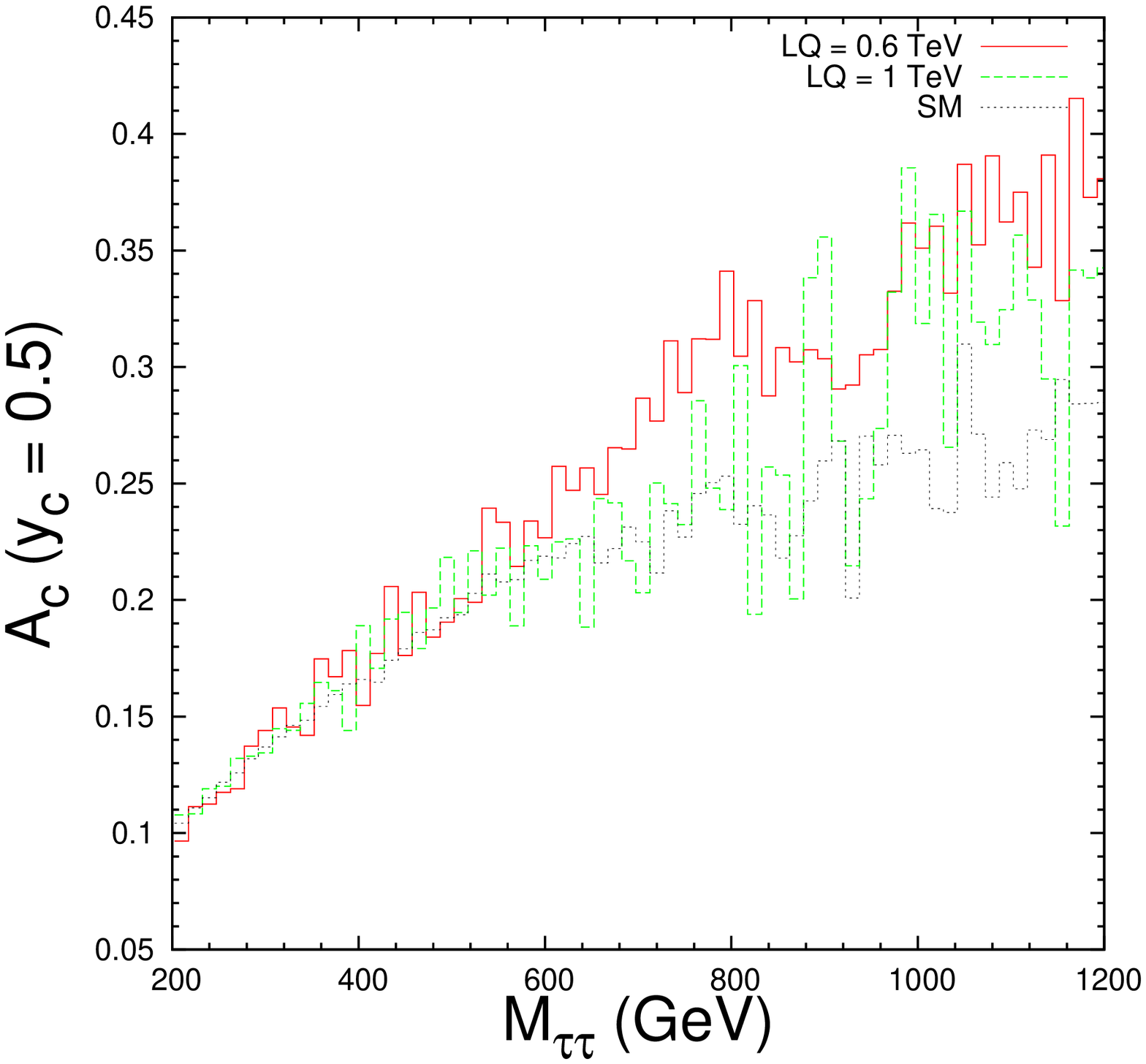}
\hskip -2.2cm
\includegraphics[angle=-90, width=.6\textwidth]{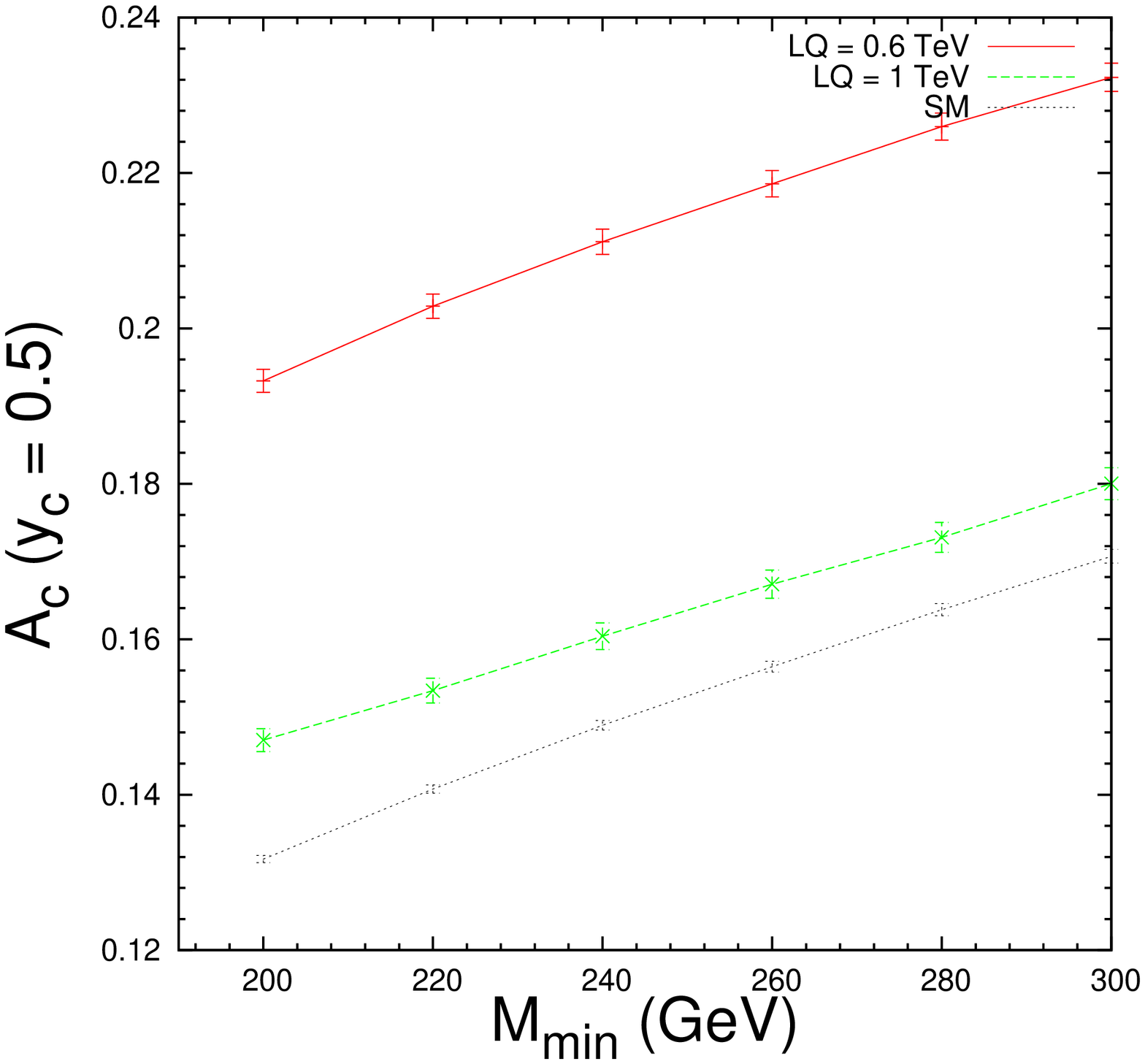}
}
\caption{\small\sf Integrated charge asymmetry for the lepto-quark model 1 (LQ-1) for $M_{LQ} = 0.6, 1$~TeV vs  $M_{\tau\tau}$ (left) and integrated over $M_{\tau\tau} \ge M_{min}$ (right).}
\label{fig:aclq1}
\end{figure}

\begin{figure}%[htb]
\centerline{
\includegraphics[angle=-90, width=.6\textwidth]{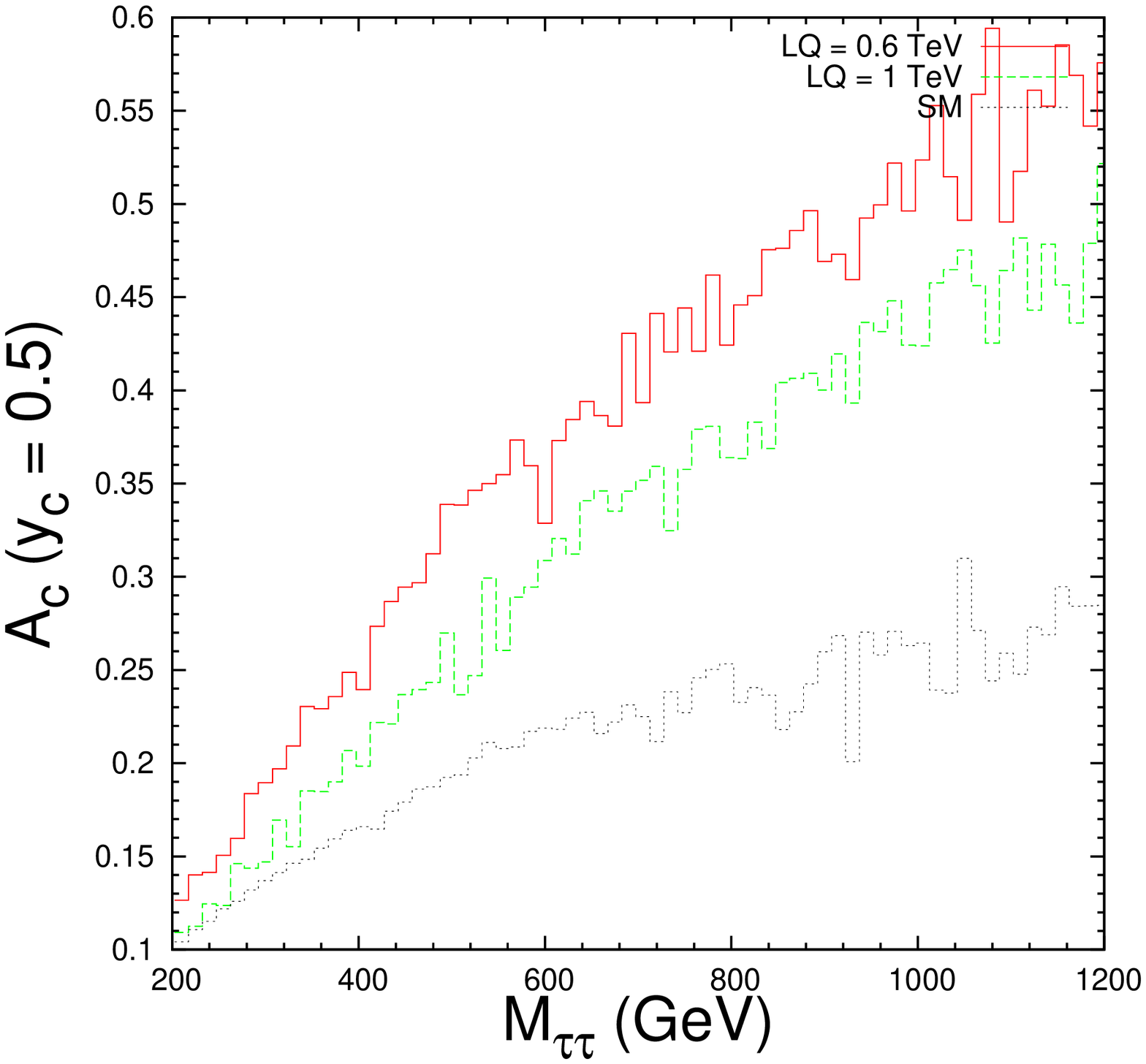}
\hskip -2.2cm
\includegraphics[angle=-90, width=.6\textwidth]{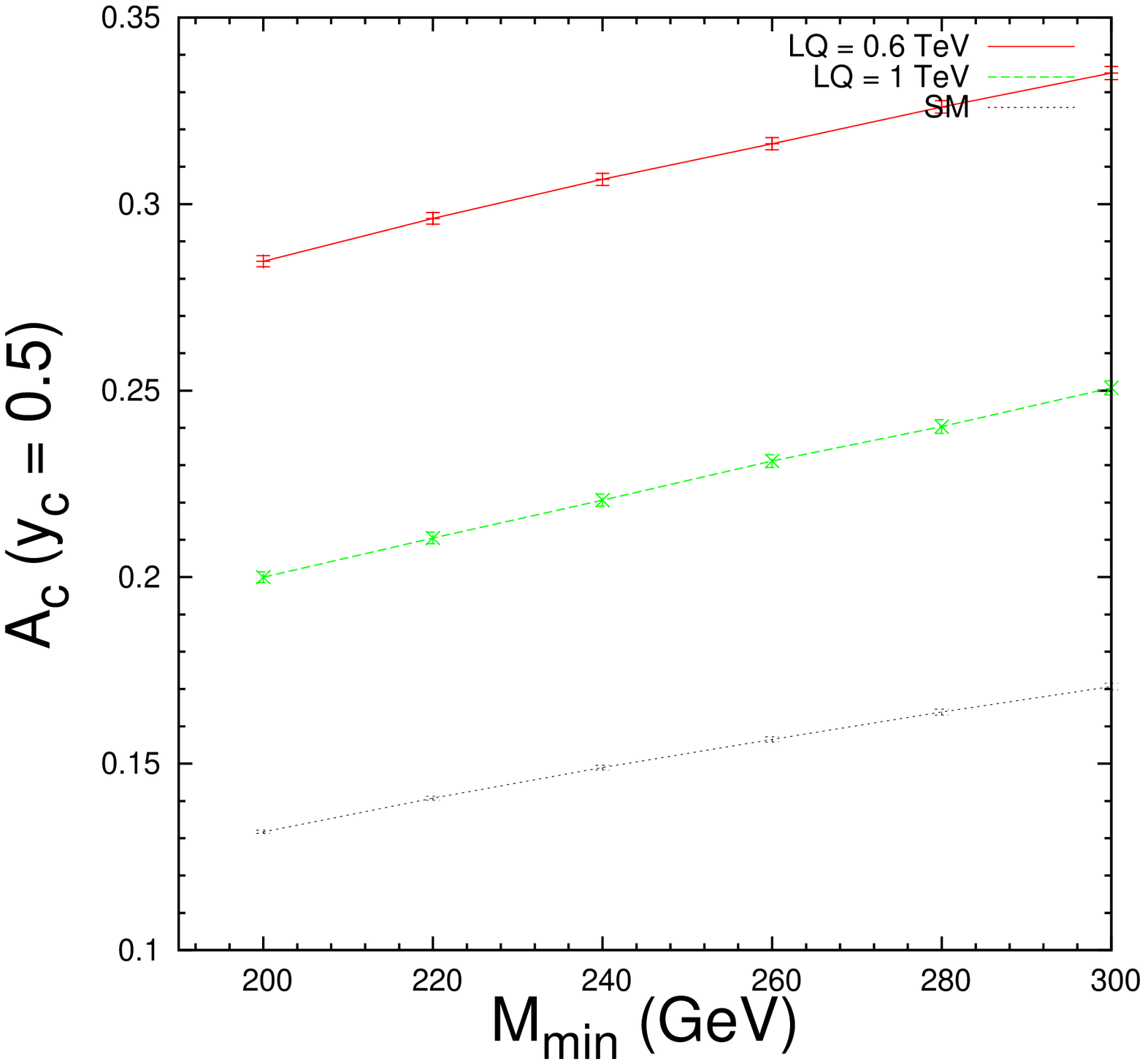}
}
\caption{\small\sf Integrated charge asymmetry for the lepto-quark model 2 (LQ-2) for $M_{LQ} = 0.6, 1$~TeV vs $M_{\tau\tau}$ (left) and  integrated over $M_{\tau\tau} \ge M_{min}$ (right).}
\label{fig:aclq2}
\end{figure}

The results for the charge asymmetry are shown in Figure~\ref{fig:aclq1} for scenario LQ-1 and in Figure~\ref{fig:aclq2} for scenario LQ-2. In both cases we show the charge asymmetry integrated over rapidity for $|y|<y_c = 0.5$ as a function of dilepton invariant mass as well as the asymmetry integrated over both rapidity and dilepton invariant mass.

%--------------------------------------------------------------------
%--------------------------------------------------------------------

%--------------------------------------------------------------------
%--------------------------------------------------------------------

In Table~\ref{t:acylq} we show the integrated $\tau$-lepton charge asymmetry in percent for two different values of the lepto-quark mass in the two models discussed above as well as for the SM. We show results for $y_c=0.5$ and for different values of $M_{min}$ with their corresponding 1$\sigma$ statistical error for an integrated luminosity of  10~fb$^{-1}$ at $\sqrt{S}=14$~TeV.

%--------------------------------------------------------------------
\begin{table}[th] 
\begin{center} 
\begin{tabular}{|c|r@{ $\pm$ }l|r@{$\pm$}l|r@{$\pm$}l|r@{$\pm$}l|r@{$\pm$}l|} 
\hline\hline
\multirow{2}{*}{Cuts }& \multicolumn{4}{|c|}{LQ-1} & \multicolumn{4}{|c|}{LQ-2} & \multicolumn{2}{|c|}{SM} 
\\\cline{2-9}
& \multicolumn{2}{|c|}{$M_{V_{2/3}} = 0.6$} & \multicolumn{2}{|c|}{$M_{V_{2/3}}= 1$} &
\multicolumn{2}{|c|}{$M_{V_{5/3}} = 0.6$} & \multicolumn{2}{|c|}{$M_{V_{5/3}} = 1$} &
\multicolumn{2}{|c|}{} \\
\hline \hline
$M_{\tau\tau} > 0.20$ TeV& 19.3 & 0.5 & 14.7 & 0.6 & 28.5 & 0.3 & 20.0 & 0.5 & 13.2 & 0.6\\
$M_{\tau\tau} > 0.22$ TeV& 20.3 & 0.5 & 15.3 & 0.7 & 29.6 & 0.3 & 21.0 & 0.5 & 14.1 & 0.7\\
$M_{\tau\tau} > 0.24$ TeV& 21.1 & 0.5 & 16.0 & 0.7 & 30.7 & 0.3 & 22.1 & 0.5 & 14.9 & 0.8\\
$M_{\tau\tau} > 0.26$ TeV& 21.9 & 0.6 & 16.7 & 0.8 & 31.6 & 0.4 & 23.1 & 0.6 & 15.6 & 0.9\\
$M_{\tau\tau} > 0.28$ TeV& 22.6 & 0.6 & 17.3 & 0.8 & 32.6 & 0.4 & 24.0 & 0.6 & 16.4 & 1.0\\
$M_{\tau\tau} > 0.30$ TeV& 23.2 & 0.6 & 18.0 & 0.9 & 33.5 & 0.4 & 25.1 & 0.6 & 17.1 & 1.1\\
\hline \hline
\end{tabular} 
\end{center}
\caption{\small\sf Integrated lepton charge asymmetry (in percent) , ${\cal A}_c(y_c)$ for lepto-quark models 1, 2 (LQ-1, 2) and the SM with $1 \sigma$-errors for 1 year of LHC data (at $\int {\cal L} dt =$ 10 fb$^{-1}$ per year). All numbers are for $y_c = 0.5$, and all masses are given in TeV. }
\label{t:acylq}
\end{table} 
%--------------------------------------------------------------------

\section{Analysis at the $\tau$-lepton decay level}

In the previous section we have identified a couple of models that produce $\tau$-lepton charge asymmetries that can differ from the SM by  factors of roughly two. In this section we study how the asymmetries are affected when the $\tau$-lepton decay is taken into consideration. For this purpose it will suffice to compare the SM asymmetry with the one induced by the LQ-2 model with $M_{V_{5/3}}=0.6$~TeV (the case resulting in the largest asymmetry in the analysis of the previous section).

We will consider the following $\tau$-lepton decay modes:

\begin{itemize}

\item Same type dilepton mode. In this case both $\tau$-leptons in the  pair undergo leptonic decay into muons or electrons:  $pp\to \tau^+\tau^- \to \ell^+ \ell^- \slashed{E}_T$, $\ell =\mu,e$. The final state is thus a muon or electron pair plus missing transverse energy, $\slashed{E}_T$ due to invisible neutrinos. Since the $\tau$-leptons are highly boosted, their decay products travel in essentially the same direction as the parent $\tau$-lepton in the lab frame. The asymmetry is therefore constructed using the direction of the muons (or electrons) instead of the $\tau$-leptons. 

The background for this mode has two origins: the $\ell^+\ell^-$ pair can arise from $t\bar{t}$, $W^+ W^-$ or $ZZ$ production, as was already the case with the analysis at the $\tau$-lepton level; or it can arise from direct Drell-Yan production of $\ell^+\ell^-$. The only handle we have on this direct background is a requirement of missing energy.

The available statistics in these two modes are decreased with respect to the number of $\tau$ pair events by about 6\%: $B(\tau\to\ell\nu\bar\nu)^2\approx 0.03$.

\item Different type dilepton mode, or events with one muon and one electron:  $pp\to \tau^+\tau^- \to \ell^+ \ell^{\prime -} \slashed{E}_T$, $\ell, \ell^\prime=\mu,e$, $\ell \neq \ell^\prime$. These modes are the cleanest from the background perspective, suffering only from the same background already present in the $\tau$-lepton level analysis: 
$t\bar{t}$, $W^+ W^-$, or $ZZ$ production. The available statistics in these modes is therefore about 6\% of the available $\tau$-lepton pairs.

\item Modes in which one $\tau$-lepton decays leptonically to a muon or an electron and the other one hadronically. In the hadronic decays of highly boosted $\tau$-leptons the resulting jet ($j_\tau$) is also approximately collinear with the original $\tau$-lepton. The asymmetry is thus constructed using the direction of the charged lepton (muon or electron) and the $j_\tau$; with the sign of the corresponding $\tau$-lepton charges being tagged by the lepton charge. The background for these modes consists of the same $t\bar{t}$, $W^+ W^-$, or $ZZ$ production present in the previous modes and in addition of $W$ plus one jet ($Wj$) production. We will limit our analysis to the two hadronic decay modes $\tau^\pm \to \pi^\pm \nu$ and $\tau^\pm \to \rho^\pm\nu$ which together account for a branching ratio near 36\%. The available statistics in these modes is therefore around 25\% of the available $\tau$-lepton pairs. 

For the $Wj$ background that meets our selection cuts, we assume a probability of 0.3\% that the QCD jet will fake a $\tau$ jet and reduce these events accordingly. This number is taken from studies by ATLAS and CMS \cite{fakejet}.

\end{itemize}

In all cases we generate the signal and background events using  {\tt MadGraph} \cite{madgraph} and implement the $\tau$-lepton decay modes with the package {\tt DECAY} \cite{decay}.  We use the following basic cuts on the 
leptons and the jets: $p_T > 6$ GeV, $|\eta| < 2.5$ and $\Delta R_{ik} < 0.4$, where the indices $i, k = \ell, j$.

\subsection{Results for dilepton modes}

\begin{figure}[htb]
\centerline{
\includegraphics[angle=-90, width=.6\textwidth]{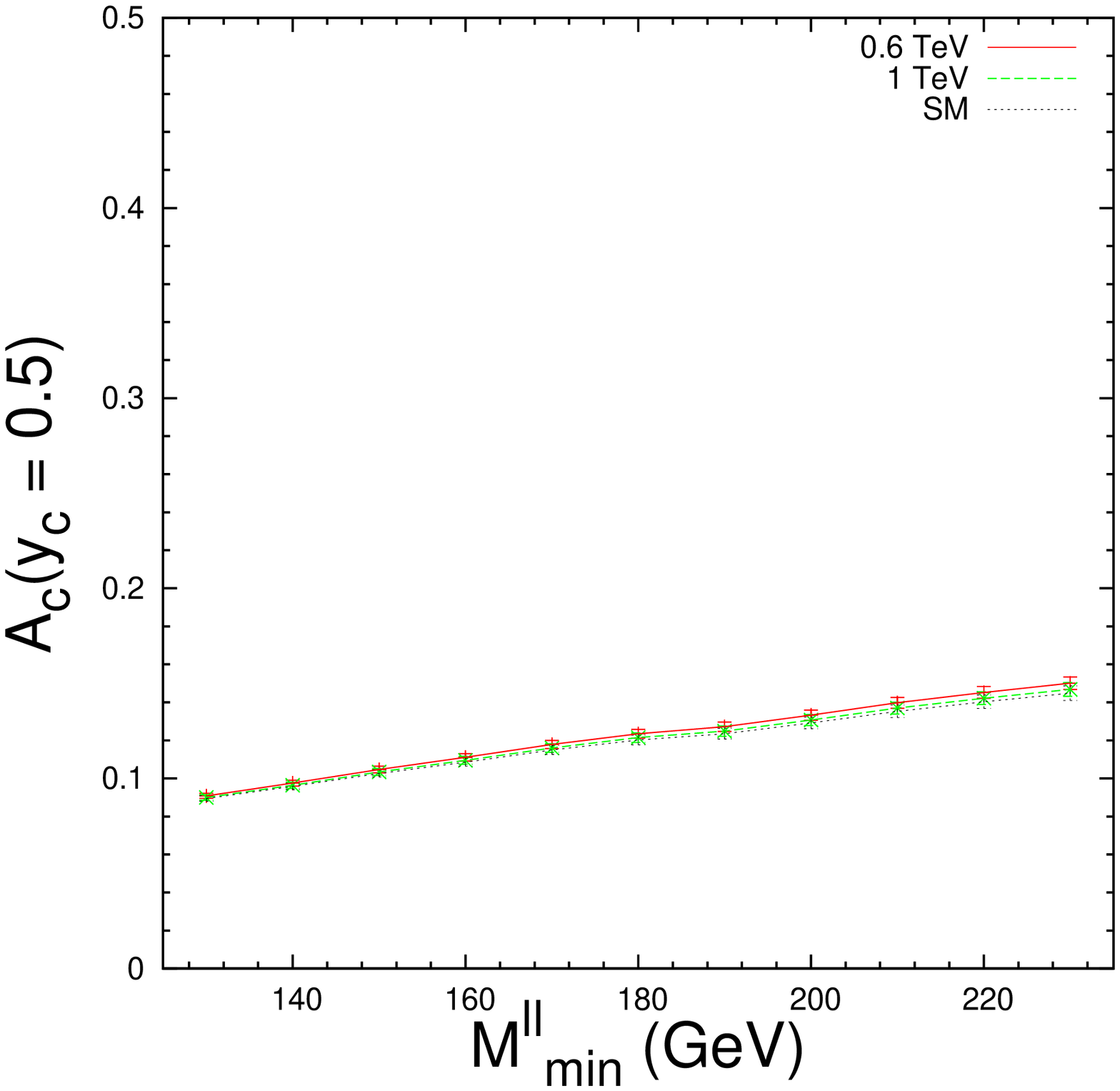}
\includegraphics[angle=-90, width=.6\textwidth]{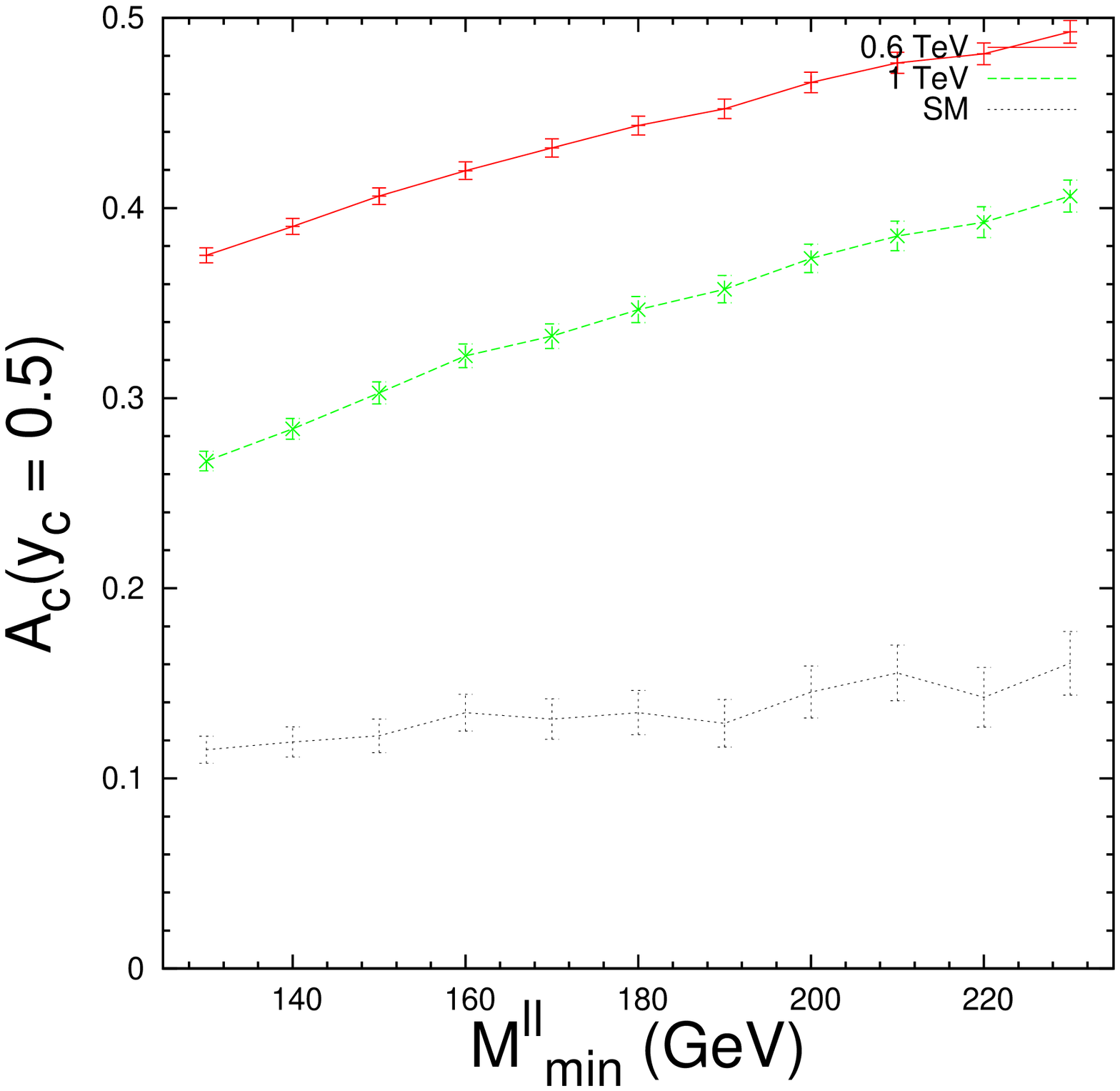}
}
\caption{\small\sf Charge asymmetry in the dilepton mode {\bf with same lepton flavor pair} for the SM and the LQ-2 model with two different lepto-quark masses as a function of a minimal $M_{\ell\ell}$ cut  without (left) and  with (right) an additional requirement of missing $E_T$ as described in the text.}
\label{fig:muac}
\end{figure}

In Figure~\ref{fig:muac} we show the charge asymmetry in the dilepton (same type) channel for both the LQ-2 model with two values of lepto-quark mass and for the SM. The curve on the left shows that the direct Drell-Yan background overwhelms the signal in this case, whereas the curve on the right shows the improvement achieved by requiring missing energy, $\slashed{E}_T > 10$~GeV in the event. We show our results for the dimuon mode, but at our level of analysis the results for $e^+e^-$ are identical. The asymmetry is shown for $y_c=0.5$ as discussed in the previous section. The asymmetry is integrated over the invariant mass of the dimuon pair with a minimal value that replaces the mimimal value of $M_{\tau\tau}$ in the previous section. We show a range for $M_{\ell\ell}$ from 130~GeV to 230~GeV. The lower end of this range is chosen to remove the $Z$ resonance where 
the SM cross-section peaks and the upper end is simply chosen for illustration.

%--------------------------------------------------------------------
%--------------------------------------------------------------------

\begin{figure}[htb]
\centerline{
\includegraphics[angle=-90, width=.6\textwidth]{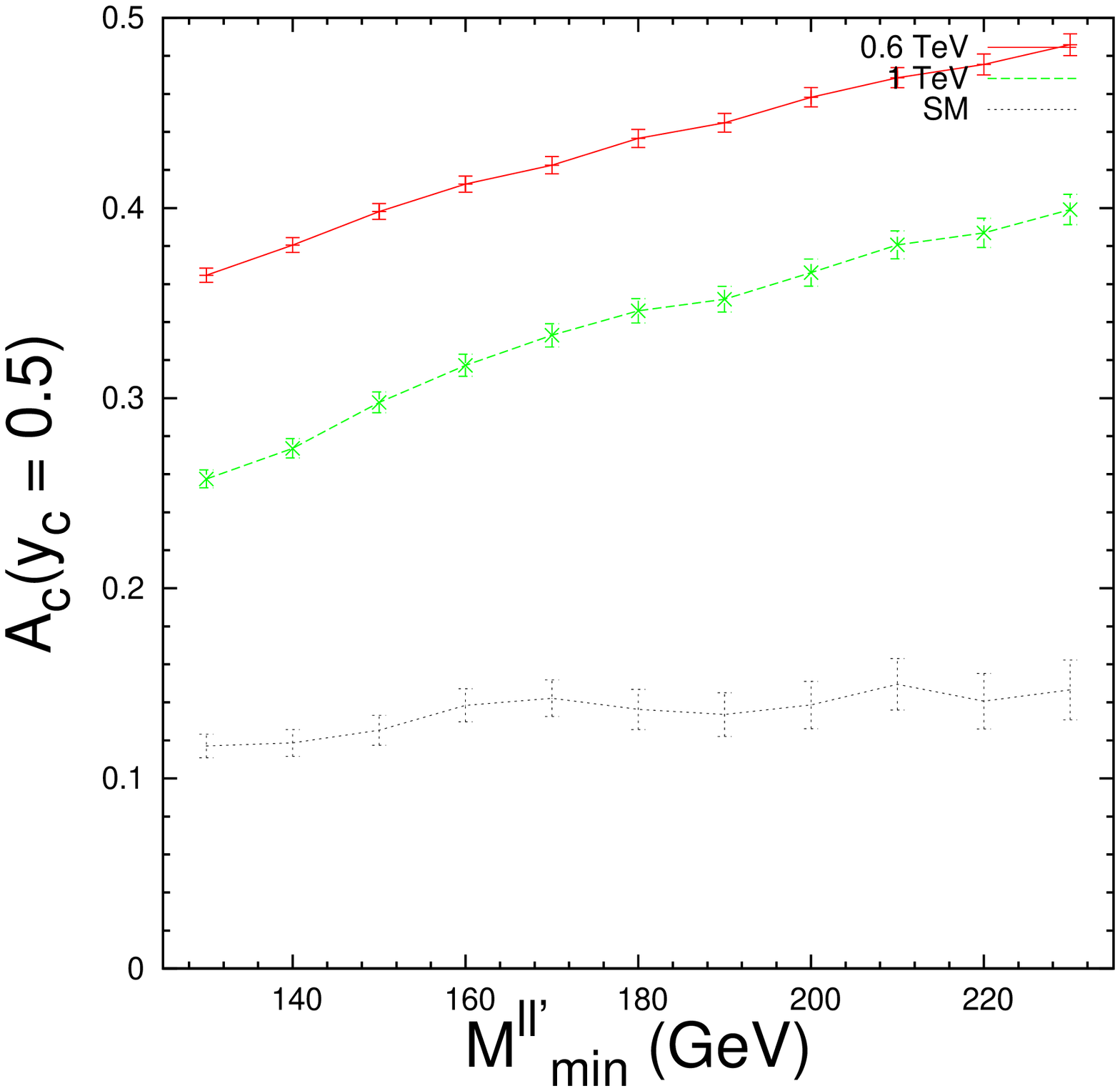}
\includegraphics[angle=-90, width=.6\textwidth]{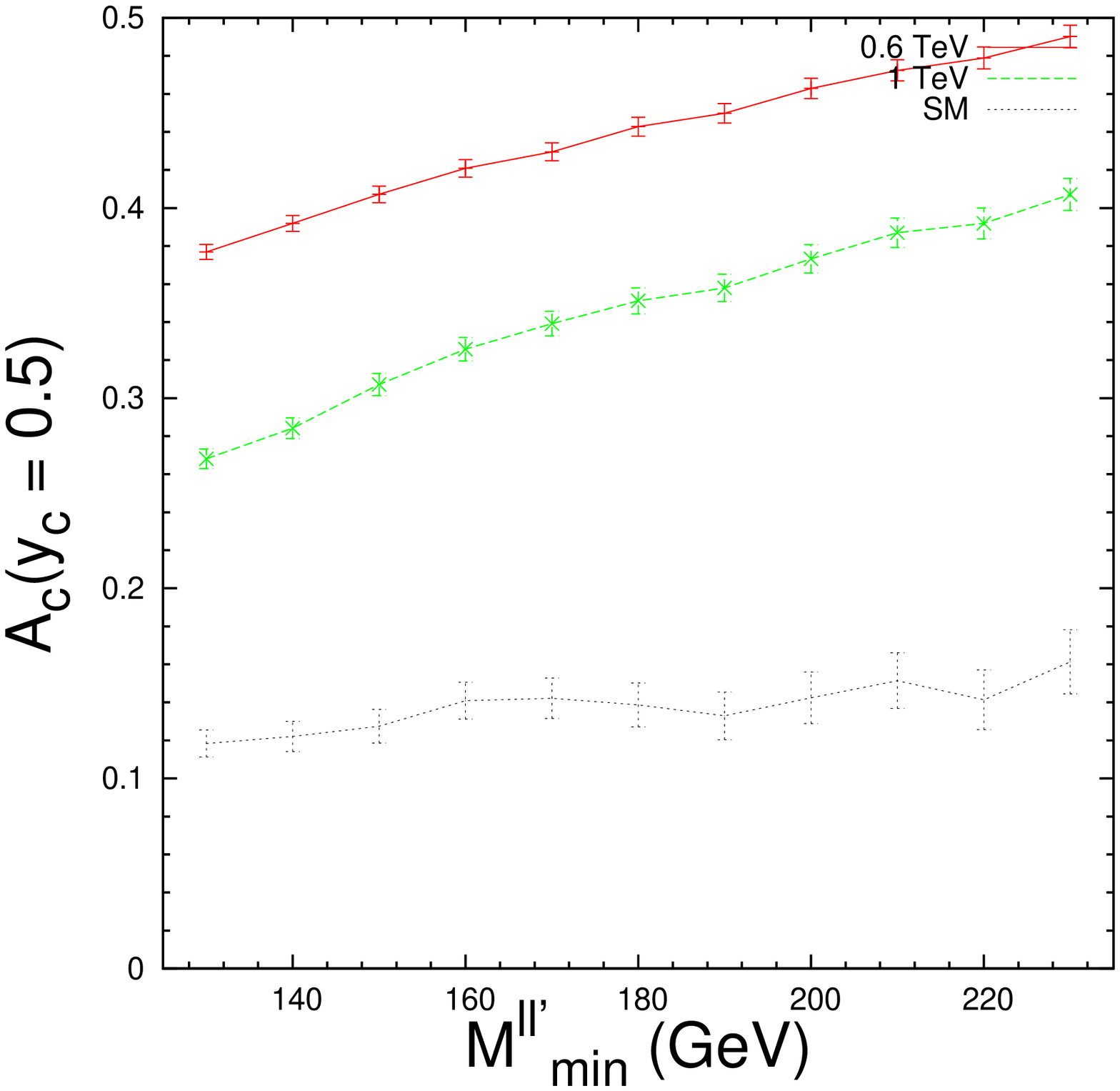}
}
\caption{\small\sf Charge asymmetry in the dilepton mode {\bf with different lepton flavor pair} for the SM and the LQ-2 model with two different lepto-quark masses as a function of a minimal $M_{\ell\ell^\prime}$ cut  without (left) and with (right) an additional requirement of missing $E_T$ as described in the text. In these plots, $\ell = \mu^+, \ell^\prime = e^-$.}
\label{fig:mueac}
\end{figure}

In Figure~\ref{fig:mueac} we show the charge asymmetry in the dilepton (different type) channel for both the LQ-2 model with two values of lepto-quark mass and for the SM. In this case we show  the results for $\mu^+ e^-$ and note that, at our level of analysis, the results for $\mu^- e^+$ are identical. In this case there is no direct Drell-Yan background, so the requirement of missing $E_T$  does not improve our signal to background ratio. The asymmetry is once again shown for $y_c=0.5$, integrated over the invariant mass of the dilepton pair. 

%--------------------------------------------------------------------
%--------------------------------------------------------------------

The integrated charge asymmetries for the four dilepton channels are given in Table~\ref{t:acylq2ll} without the missing $E_T$ requirement in left table and with the additional $\slashed{E}_T >10$~GeV requirement in the right table. This missing energy requirement only affects the $e^+e^-$ and $\mu^+\mu^-$ modes where it removes the direct Drell-Yan background. The errors shown in the tables are the statistical errors corresponding 10 fb$^{-1}$. These results show the usefulness of the charge asymmetry as a discriminator for new physics affecting the $\tau$-lepton.

%--------------------------------------------------------------------
\begin{table}[th]
\hspace{-3.5cm}
\begin{minipage}[b]{0.25\linewidth}\begin{center}
\begin{tabular}{|c|r@{ $\pm$ }l|r@{$\pm$}l|r@{$\pm$}l|}
\hline\hline
\multirow{2}{*}{$(\ell, \ell^\prime)$ } & \multicolumn{4}{|c|}{$LQ-2$ Model} & \multicolumn{2}{|c|}{SM}
\\\cline{2-5}
& \multicolumn{2}{|c|}{$M_{V_{5/3}} = 0.6$} & \multicolumn{2}{|c|}{$M_{V{5/3}}= 1$} &\multicolumn{2}{|c|}{} \\
\hline \hline
$(\mu^+, \mu^-)$ & 9.1  & 0.5 &  9.0 & 0.6 &  8.9 & 0.6\\
$(e^+, e^-)$     & 9.2  & 0.5 &  9.1 & 0.6 &  9.1 & 0.6\\
$(e^+, \mu^-)$   & 36.2 & 1.6 & 25.6 & 2.0 & 12.1 & 2.3\\
$(\mu^+, e^-)$   & 36.5 & 1.6 & 25.8 & 2.0 & 11.7 & 2.3\\
\hline \hline
\end{tabular}
\end{center}\end{minipage}
\hspace{3.5cm}
\begin{minipage}[b]{0.25\linewidth}\begin{center}
\begin{tabular}{|c|r@{ $\pm$ }l|r@{$\pm$}l|r@{$\pm$}l|}
\hline\hline
\multirow{2}{*}{$(\ell, \ell^\prime)$ } & \multicolumn{4}{|c|}{$LQ-2$ Model} & \multicolumn{2}{|c|}{SM}
\\\cline{2-5}
& \multicolumn{2}{|c|}{$M_{V_{5/3}} = 0.6$} & \multicolumn{2}{|c|}{$M_{V{5/3}}= 1$} &\multicolumn{2}{|c|}{} \\
\hline \hline
$(\mu^+, \mu^-)$ & 37.5 & 1.7 & 26.7 & 2.2 & 11.5 & 2.4\\
$(e^+, e^-)$     & 37.7 & 1.7 & 26.6 & 2.2 & 12.4 & 2.4\\
$(e^+, \mu^-)$   & 37.4 & 1.7 & 26.6 & 2.2 & 12.2 & 2.4\\
$(\mu^+, e^-)$   & 37.7 & 1.7 & 26.8 & 2.2 & 11.8 & 2.4\\
\hline \hline
\end{tabular}
\end{center}\end{minipage}
\caption{\small\sf Integrated lepton charge asymmetry ${\cal A}_{\ell\ell^\prime}$ (in percent) for the
$LQ-2$ model and the SM with $M^{\ell\ell^\prime}_{min} = 130$ GeV (a) without (Left table) and (b) with the requirement of 
a minimum missing $E_T$ (Right table). The $1 \sigma$-errors correspond to statistics for
one year of LHC data (at $\int {\cal L} dt =$ 10 fb$^{-1}$ per year). All the masses are given in TeV.
}
\label{t:acylq2ll}
\end{table}
%--------------------------------------------------------------------

\subsection{Results for the $\ell j_\tau$ modes}

\begin{figure}[htb]
\centerline{
\includegraphics[angle=-90, width=.6\textwidth]{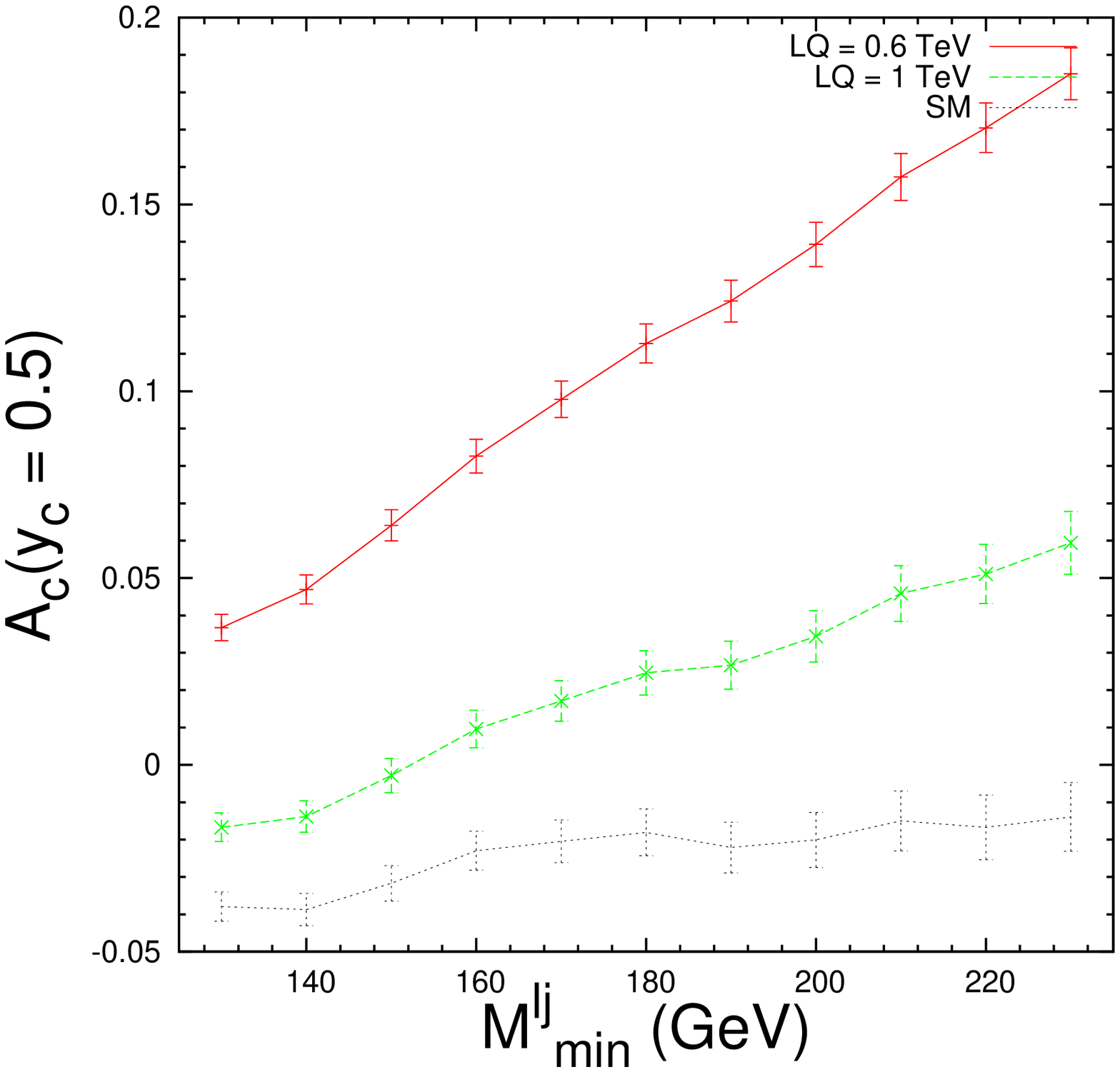}
\includegraphics[angle=-90, width=.6\textwidth]{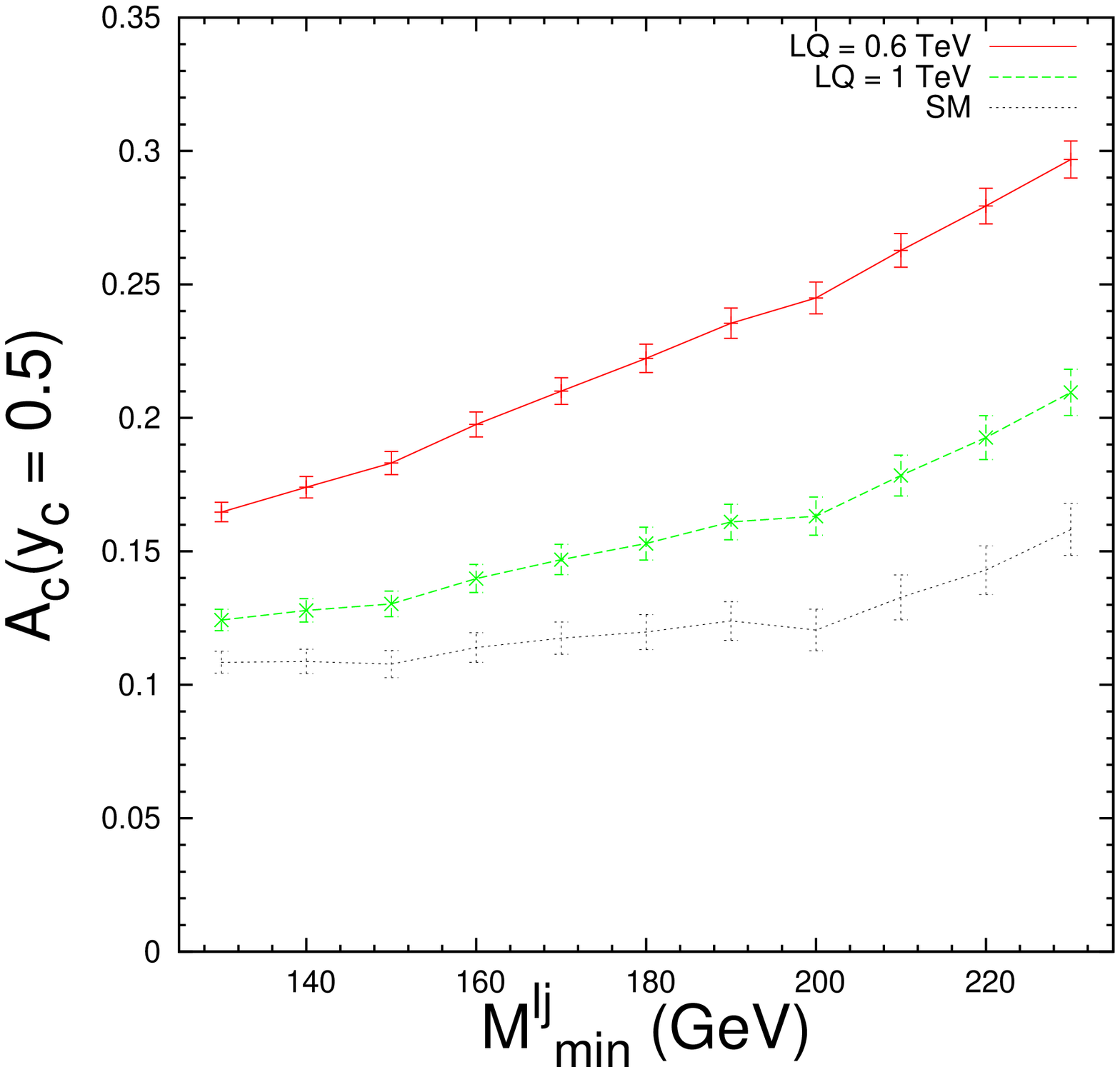}
}
\caption{\small\sf Charge asymmetry $A_{lj}$ with a missing $E_T$ requirement. The results for $\ell^+ j$ and $\ell^- j$ are shown respectively in the left-side and right-side plots. In both cases $\ell^\pm = e^\pm, \mu^\pm$ and the jet is discussed in the text.}
\label{fig:lpjet}
\end{figure}

These modes suffer from the additional $Wj$ background in which the QCD jet fakes a $\tau$ jet. This background is very significant even after we reduce it by assuming a 0.3\% misidentification rate and this is evident in the difference between the results for $\ell^+ j$ and $\ell^- j$ in Figures~\ref{fig:lpjet} between left and right figures. We can understand the difference between these asymmetries qualitatively because the $Wj$ background events themselves have an asymmetry that is different for $W^+j$ and $W^-j$, $-6.3\%$ and $11\%$ respectively with our cuts. In addition, the background is larger for $\ell^+ j$ since  $\sigma(W^+ j)>\sigma(W^-j)$ at $pp$ colliders.

%%%%%%%%% Lepton-jet plots and tables
%--------------------------------------------------------------------
%--------------------------------------------------------------------

The results for the integrated asymmetries in these two cases are presented in Table~\ref{t:acylq2lj_et} requiring the events to have $\slashed{E}_T >10$~GeV. The missing $E_T$ requirement is not very important in this case.

%--------------------------------------------------------------------
\begin{table}[th]
\begin{center}
\begin{tabular}{|c|r@{ $\pm$ }l|r@{$\pm$}l|r@{$\pm$}l|}
\hline\hline
\multirow{2}{*}{} & \multicolumn{4}{|c|}{$LQ-2$ Model} & \multicolumn{2}{|c|}{SM}
\\\cline{2-5}
& \multicolumn{2}{|c|}{$M_{V_{5/3}} = 0.6$} & \multicolumn{2}{|c|}{$M_{V{5/3}}= 1$} &\multicolumn{2}{|c|}{} \\
\hline \hline
$(\ell^+, j)$ & 3.7  & 0.4 & -1.7 & 0.4 & -3.7 & 0.4\\
$(j, \ell^-)$ & 16.5  & 0.4 & 12.4 & 0.4 & 10.9 & 0.4\\
\hline \hline
\end{tabular}
\end{center}
\caption{\small\sf Integrated lepton charge asymmetry ${\cal A}_{\ell j}$ (in percent) for the
$LQ-2$ model and the SM with $M^{\ell j}_{min} = 130$ GeV with the missing $E_T$ requirement. The $1 \sigma$-errors correspond to statistics for
one year of LHC data (at $\int {\cal L} dt =$ 10 fb$^{-1}$ per year).}
\label{t:acylq2lj_et}
\end{table}
%--------------------------------------------------------------------

These results show that the $\ell j_\tau$ channel is less promising for using the charge asymmetry as a discriminator of new physics, but that it can still serve to distinguish some models. 

Interestingly, the asymmetries in the $\ell^\pm j$ channels, with the $W^\pm j$ background removed at the Monte Carlo level, are not the same as those in the $\ell^+\ell^{\prime -}$ channels. This is due to the different effect of requiring the same minimum invariant mass for $\ell \ell^\prime$ and $\ell j$, with the cut in  $\ell \ell^\prime$ being more efficient at rejecting SM background. The difference is simply due to the different kinematics for two-body decay vs three-body decay.  As a consequence of this difference it would be difficult to combine the different channels to improve the  statistical error even if the $W^\pm j$ background could be controlled effectively.

\section{Summary and Conclusion}

The forward-backward asymmetry has been a very useful tool to obtain information about the SM couplings of fermions to the $Z$ boson. At the LHC it will not always be possible to reconstruct a forward-backward asymmetry, but some of the same information can be obtained from charge asymmetries. In this paper we have emphasized that the charge asymmetry can also play an important role in the search for new physics in $\tau$-pair production at the LHC.

We have investigated some kinematic properties of this charge asymmetry concluding that a value of $y_c$ near 0.5 is optimal in this case. The asymmetry can be constructed as a function of lepton pair invariant mass, $M_{\tau\tau}$, but a better probe of new physics is obtained by integrating the charge asymmetry over the available $M_{\tau\tau}$ range with a minimum cut, $M_{min}$. Originally this cut serves the purpose of removing the SM $Z$ background. We found that the integrated asymmetry increases by increasing $M_{min}$ far above the $M_Z$, although at the cost of lost statistics. We have explored the optimal value of $M_{min}$ in the context of two examples. 

We have illustrated two generic scenarios of new physics models that single out the $\tau$-lepton and would not show up in $\mu^+\mu^-$ or $e^+e^-$ pair production. We discuss a non-universal $Z^\prime$ as an example of a new resonance that prefers to decay into tau-leptons.  We also considered certain lepto-quarks as an example of non-resonant new physics. They single out the tau-lepton by associating the third generation leptons with the first generation quarks. We have illustrated examples of new physics that can induce $\tau$ charge asymmetries that are easy to distinguish from the SM. 

We have first presented an analysis of the charge asymmetry at the $\tau^+\tau^-$ level based on the premise that $\tau^+\tau^-$ samples have already been identified by the experiments. To back up our conclusions we have also studied selected $\tau$ decays with their associated backgrounds. We find that the dilepton modes are very clean to study the charge asymmetry. The lepton plus jet modes are less sensitive, but can still provide some discriminating power in the charge asymmetry.

In conclusion we find that an experimental study of the $\tau$-lepton charge asymmetry at the LHC can provide valuable information in the search for new physics. Clearly a detector level study by the experimental collaborations will be necessary to accurately quantify the possibilities of this proposal, and we encourage both ATLAS and CMS to conduct such studies.

\begin{acknowledgments}

This work was supported in part by DOE under contract number
DE-FG02-01ER41155. We thank David Atwood for useful discussions.

\end{acknowledgments}

\end{document}